\documentclass[10pt]{article}
\usepackage[utf8]{inputenc}

\usepackage{cite}
\usepackage{amsmath,amssymb,amsfonts}
\usepackage{algorithmic}
\usepackage[left=1.62cm,right=1.62cm,top=1.9cm]{geometry}
\usepackage{graphicx}
\usepackage{textcomp}
\usepackage{xcolor}
\def\BibTeX{{\rm B\kern-.05em{\sc i\kern-.025em b}\kern-.08em
    T\kern-.1667em\lower.7ex\hbox{E}\kern-.125emX}}

\usepackage{tikz}
\usepackage{braket}
\usetikzlibrary{quantikz}
\usepackage{tabularx,booktabs,multirow}
\usepackage{mathtools}
\usepackage{marvosym}
\usepackage[ruled,linesnumbered]{algorithm2e}
\usepackage{hyperref}
\usepackage{url}
\newcommand{\etal}{{\em et al.}}
\newcommand{\ie}{{\em i.e.}}

\begin{document}

\title{Graph Neural Networks for Parameterized Quantum Circuits \\Expressibility Estimation}

\author{Shamminuj Aktar$^{1}$,  Andreas Bärtschi$^{2}$, Diane Oyen$^{2}$, Stephan Eidenbenz$^{2}$, Abdel-Hameed A. Badawy$^{1}$}

\date{
    \small
    $^1$ Klipsch School of Electrical and Computer Engineering, New Mexico State University\\
    $^2$ CCS-3 Information Sciences, Los Alamos National Laboratory\\
    Email: saktar@nmsu.edu, baertschi@lanl.gov, doyen@lanl.gov, eidenben@lanl.gov,badawy@nmsu.edu
}

\maketitle

\begin{abstract}
Parameterized quantum circuits (PQCs) are fundamental to quantum machine learning (QML), quantum optimization, and variational quantum algorithms (VQAs). The expressibility of PQCs is a measure that determines their capability to harness the full potential of the quantum state space. It is thus a crucial guidepost to know when selecting a particular PQC ansatz. However, the existing technique for expressibility computation through statistical estimation requires a large number of samples, which poses significant challenges due to time and computational resource constraints. This paper introduces a novel approach for expressibility estimation of PQCs using Graph Neural Networks (GNNs). We demonstrate the predictive power of our GNN model with a dataset consisting of 25,000 samples from the noiseless IBM QASM Simulator and 12,000 samples from three distinct noisy quantum backends. The model accurately estimates expressibility, with root mean square errors (RMSE) of 0.05 and 0.06 for the noiseless and noisy backends, respectively. We compare our model’s predictions with reference circuits from~Sim~\etal~and IBM Qiskit’s hardware-efficient ansatz sets to further evaluate our model’s performance. Our experimental evaluation in noiseless and noisy scenarios reveals a close alignment with ground truth expressibility values, highlighting the model’s efficacy. Moreover, our model exhibits promising extrapolation capabilities, predicting expressibility values with low RMSE for out-of-range qubit circuits trained solely on only up to 5-qubit circuit sets. This work thus provides a reliable means of efficiently evaluating the expressibility of diverse PQCs on noiseless simulators and hardware. 
\end{abstract}


\section{Introduction}
Quantum computing can potentially transform a wide range of industries by solving computational problems that are classically intractable. However, in the current era of noisy intermediate-scale quantum (NISQ) computing~\cite{preskill2018quantum}, quantum devices face limitations regarding the number of qubits, coherence time, and noise. These constraints present significant challenges to harnessing the full potential of quantum computing. Among quantum algorithms, variational quantum algorithms (VQAs), which belong to the category of hybrid quantum-classical (HQC) algorithms, are widely viewed as promising candidates well-suited to the limitations imposed by NISQ devices~\cite{cerezo2021variational,endo2021hybrid}. In VQA, a problem-specific cost function is evaluated on a quantum computer by executing and measuring a parameterized quantum circuit (PQC), and concurrently, a classical optimizer iteratively adjusts the parameters of the parameterized quantum circuit to minimize this cost. Similarly, other approaches such as the variational quantum eigensolver (VQE)~\cite{peruzzo2014variational}, the quantum approximate optimization algorithm (QAOA)~\cite{farhi2014quantum}, quantum autoencoders (QAE)~\cite{romero2017quantum}, or quantum neural networks (QNN)~\cite{jeswal2019recent} leverage quantum subroutines to generate parameterized trial states. These parameters are optimized to approach optimal or near-optimal solutions for the corresponding objective function. 

PQCs are thus the (quantum) heart of HQC approaches. The structure of a PQC in terms of one-qubit and two-qubit gates is called an ansatz. Varying the parameter values within the ansatz circuit would allow for a systematic exploration of the solution space of a target problem. Therefore, the performance of HQC algorithms critically depends on the exploration ability of the chosen PQC. However, designing an effective PQC with low circuit depth and a concise parameter space that accurately represents the solution space remains a significant challenge.

The importance of PQCs has spurred the development of various ansatz designs tailored to specific quantum algorithms and hardware platforms. These include problem-specific PQCs customized for particular applications~\cite{saib2021effect, moll2018quantum}, as well as hardware-efficient PQCs optimized for execution on NISQ devices~\cite{kandala2017hardware}. These diverse ansatz designs aim to strike a balance between computational complexity and \emph{expressibility}, ensuring efficient exploration of solution spaces within the limitations of quantum hardware.

% However, designing an effective PQC with low circuit depth and a small number of parameters accurately representing the solution space is challenging. The importance of PQCs has led to the development of various ansatz designs, % for various quantum algorithms, 
% including problem-specific PQCs~\cite{saib2021effect, moll2018quantum} and hardware-efficient PQCs~\cite{kandala2017hardware}. 

The ability of PQCs to generate or explore states in the Hilbert space is called the expressibility of the PQC~\cite{sim2019expressibility}. This measure is defined as the Kullback-Leibler (KL) divergence~\cite{joyce2011kullback} between the fidelity distribution estimated from PQCs and the distribution resulting from a Haar-random unitary~\cite{zyczkowski2005average}. The estimated fidelity distribution is computed by measuring the overlap between two quantum states prepared with different sets of parameters. However, directly measuring expressibility poses challenges due to the requirement for many fidelity samples. For instance, experiments in~Sim~\etal~\cite{sim2019expressibility} utilized 5,000 fidelity samples for four-qubit circuits. Moreover, the influence of specific noise characteristics inherent in quantum devices affects the performance of PQCs within a VQA algorithm~\cite{saib2021effect}.

Recently, there has been a growing trend of using machine learning (ML) to address complex issues in quantum systems. Specifically, researchers are exploring graph-based learning techniques, viewing quantum circuits as graphs. This approach utilizes the inherent structure of quantum circuits, where gates are nodes and connections are edges. For instance, some recent works have delved into reliability estimation of quantum circuits~\cite{Wang2022torchQuantum,saravanan2022data,alrahisGNN4REL2022}. To date, no prior research has focused specifically on directly predicting the expressibility of PQCs.

\begin{figure*}[t!]
    \centering
    \includegraphics[width=\linewidth]{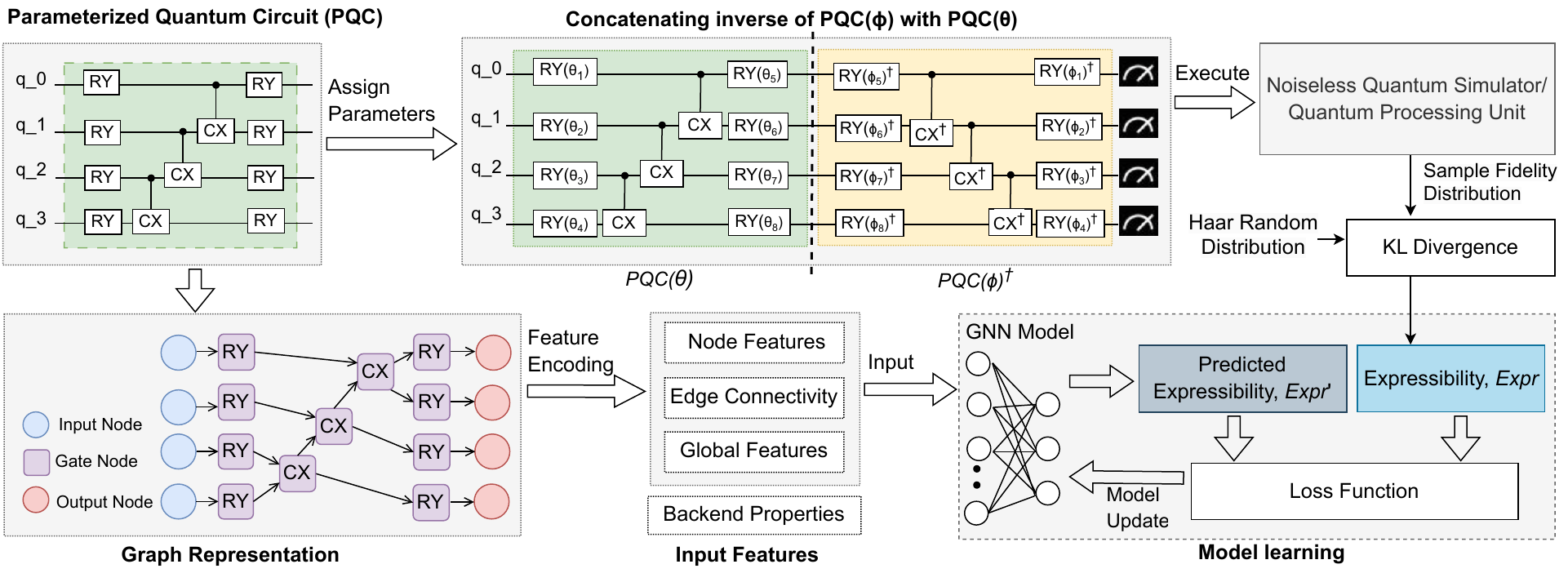}
    \caption{Our proposed framework for predicting expressibility of Parameterized Quantum Circuits (PQCs) using Graph Neural Networks (GNNs):\newline First, we derive a graph representation of a PQC with input, gate, and output nodes encoded with node features, edge connectivity, global circuit features, and target backend properties. Its true expressibility value $Expr$ is computed using Equation~\eqref{eq:expr}. %where sample fidelity distribution and Haar random distribution are computed as described in Section~\ref{sec:expr-calc}. 
    In the model learning phase, the GNN model takes the encoded features as inputs and adjusts its parameters to minimize the difference between predicted and true expressibility values.}
    \label{fig:diagram}
\end{figure*}

To address these challenges, we propose a novel approach that employs a Graph Neural Network (GNN)~\cite{scarselli2008graph} to predict the expressibility of PQCs in a specific quantum device (see Figure~\ref{fig:diagram}). We adapt the graph transformer model from Wang et al.~\cite{Wang2022torchQuantum}, modifying it to suit our specific task requirements. In our proposed approach, the GNN model leverages the inherent structural information in the connectivity patterns of the PQCs to learn the relationship between PQCs and their expressibility. First, we collect a large dataset containing randomly generated PQCs and their expressibility values sourced from different quantum backends. Then, we extract graph representations from these circuits to generate node features and edge connectivity and incorporate global circuit features and backend noise information. The model learns from these features and adjusts its parameters to minimize the difference between the predicted and the actual target expressibility values. Evaluation of a comprehensive dataset containing samples from both noiseless and noisy backends demonstrated the high predictive power of our trained model. Since there is no prior work on expressibility prediction using ML techniques, we conduct a comparison between the predicted expressibility values from our model and the original expressibility values from reference circuits used in~\cite{sim2019expressibility} and hardware-efficient ansatz sets from IBM Qiskit~\cite{realAmplitude}. This comparison validates the performance of the trained model on the PQCs used in different application domains. 
\vspace{1ex}

The contributions of the paper are as follows:
\begin{itemize}
    \item We collected a comprehensive dataset of PQCs and their associated expressibility values, comprising 25,000 samples from the noiseless quantum backend and 12,000 samples from noisy quantum backends for training and evaluation. The training/testing set covers PQCs up to `8'~qubits while the extrapolation set ranges up to `10'~qubits.
    
    \item We train a graph transformer model using the comprehensive dataset to predict expressibility from input samples directly. Our trained model showcases significant prediction accuracy, achieving root mean square errors (RMSE) of $0.05$ and $0.06$ for the noiseless and noisy backends, respectively.
    
    \item To further assess our trained model, we undertake a comparative evaluation using 19 reference circuits from~\cite{sim2019expressibility} and IBM Qiskit’s hardware-efficient ansatz sets~\cite{realAmplitude}. The evaluation results demonstrate close alignment with the actual expressibility values for the evaluation PQC sets on the noiseless and noisy backends. 
    
    \item On the noiseless backend, the reference circuits and hardware-efficient ansatz sets of IBM Qiskit achieve RMSE values of $0.05$ and $0.06$, respectively. However, specific device characteristics primarily influence the accuracy on noisy backends, resulting in average RMSE values of $0.07$ and $0.08$, respectively.

    \item Our proposed model provides accurate expressibility approximations even for extrapolated (out-of-range) circuits. 
\end{itemize}

The rest of the paper is organized as follows: Section~\ref{sec:background} provides background on VQAs, PQCs, and expressibility. Section~\ref{sec:framework} includes our proposed approach for expressibility prediction. Section~\ref{sec:exp} discusses experimental model evaluation results for noisy and noiseless samples. Section~\ref{sec:related} reviews related works on the applications of ML techniques in quantum computing. Finally, Section~\ref{sec:conclusion} concludes the paper.

\section{Background}

%\section{Background}
\label{sec:background}
\subsection{Variational Quantum Algorithms}
%%%Add references
Variational Quantum Algorithms (VQAs) have become one of the leading candidates in achieving application-oriented quantum computational advantage on NISQ devices~\cite{cerezo2021variational}. Their hybrid quantum-classical approach shows promising potential for noise resilience, making them particularly well-suited for practical applications in quantum computing. A VQA includes a PQC for generating trial quantum states, a problem-specific cost function, quantum measurements to evaluate the cost function on the trial states, and a classical optimizer adjusting PQC parameters based on measurement results. The choice of the cost function, derived from the Hamiltonian of the problem under investigation, affects the performance of VQAs~\cite{cerezo2021cost}. Moreover, the structure of the PQC greatly influences VQA performance, impacting convergence speed and solution accuracy. Some common PQCs used in VQAs are problem-inspired ansatzes~\cite{farhi2014quantum,cerezo2021variational}, hardware-efficient ansatzes~\cite{kandala2017hardware} and variable-structure ansatzes~\cite{grimsley2019adaptive}.
Typical applications of VQAs span various application domains. The Variational Quantum Eigensolver (VQE)~\cite{peruzzo2014variational} is used to find ground state energies of quantum systems in quantum chemistry. The Quantum Approximate Optimization Algorithm (QAOA)~\cite{farhi2014quantum} can approximate solutions for optimization problems like Max-Cut~\cite{guerreschi2019qaoa}. Furthermore, Quantum Machine Learning (QML) leverages VQAs as quantum analogs of classical neural networks, facilitating tasks such as classification and generative modeling with enhanced efficiency and capabilities\cite{biamonte2017quantum}.
%\subsection{Quantum Machine Learning}
\subsection{Parameterized Quantum Circuits (PQCs)}
In near-term hybrid quantum-classical (HQC) algorithms, PQCs link between quantum and classical computers: quantum machines use PQCs to make predictions and approximations, while classical machines update the circuit parameters~\cite{mcclean2016theory}. Notable examples include VQE~\cite{peruzzo2014variational}, QAOA~\cite{farhi2014quantum}, among others. There is also a connection between PQCs and classical neural networks, where the changeable parts of PQCs resemble the weights and biases in classical neural networks~\cite{schuld2020circuit, killoran2019continuous}.
By definition, PQCs represent tunable unitary operations, denoted as $U_{\vec{\theta}}$, that are applied to a chosen initial state. A PQC consists of several fixed quantum gates and trainable single qubit quantum gates \ie~$R_x(\theta_i)$, $R_y(\theta_j)$, and $R_z(\theta_k)$ rotations.
Such circuits offer the flexibility to adjust their parameters to achieve specific quantum operations or optimize their performance within a quantum algorithm. PQCs are generally designed to be versatile, and their capacity to cover a broader region of the solution space enhances their expressiveness. Recently, “expressibility” has been proposed as a metric for evaluating PQCs to assess the effectiveness and versatility of PQCs in representing trial quantum states and performing quantum operations~\cite{sim2019expressibility}. 
Recent works investigate expressibility to find optimal PQCs for quantum machine-learning~\cite{chen2021expressibility, kobayashi2022overfitting, hubregtsen2021evaluation, tangpanitanon2020expressibility, mcclean2018barren} and optimization problems~\cite{holmes2022connecting, nakaji2021expressibility}.

\begin{figure*}[t!]
    \centering
    \includegraphics[width=\linewidth]{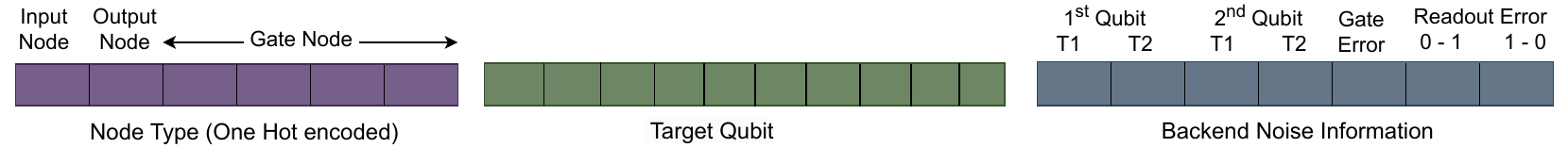}
    \vspace*{-4ex} %% accounts for graphics whitespace
    \caption{Node feature vector (23 digits): six digits for node type, ten digits for target qubit(s), and seven digits for backend calibration Information.}
    \vspace*{2ex}
    \label{fig:node-feature}
\end{figure*}

\begin{figure*}[t!]
    \centering
    \includegraphics[width=\linewidth]{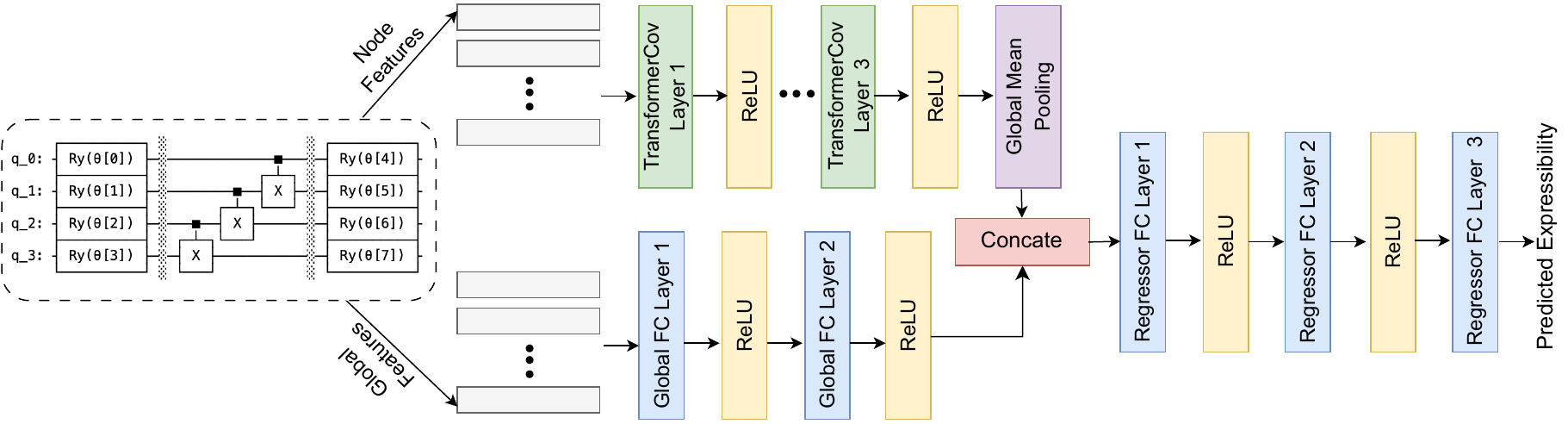}
    \caption{Overview of the GNN model for expressibility prediction of PQC. The node features are passed to the graph transformer layer to capture the neighboring correlations from nodes. Global features are propagated through fully connected layers and then concatenated with the feature vector from transformer layers. The aggregated feature vector is then directed to fully connected regressor layers to perform the regression task of expressibility prediction for PQCs.}
\label{fig:gnn}
\end{figure*}

\subsection{Expressibility}
\label{sec:expr}
In the context of PQC, \emph{expressibility} measures how effectively these circuits can produce quantum states that accurately represent the Hilbert space~\cite{sim2019expressibility}. It is defined as
\begin{align}
    \label{eq:expr}
    Expr = D_{KL} (P_{PQC}(F;\vec{\theta}) || P_{Haar}(F)),
\end{align}
where $P_{PQC}(F;\vec{\theta})$ represents the estimated probability distribution of quantum state fidelities derived from a selection of quantum states generated by the PQC over random parameters $\vec{\theta}$, and $P_{Haar}(F)$ represents the maximally expressive uniform distribution of quantum fidelities originating from an ensemble of Haar random states~\cite{zyczkowski2005average}. Thus, Expressibility is the Kullback-Leibler (KL) divergence~\cite{joyce2011kullback} of $P_{PQC}(F;\vec{\theta})$ from $P_{Haar}(F)$~\cite{sim2019expressibility}. To estimate the fidelity distribution $P_{PQC}(F;\vec{\theta})$, we uniformly sample pairs of parameter sets $\vec{\theta}$ and $\vec{\phi}$ for the PQC on each iteration, resulting in the generation of random quantum states $\ket{\psi_{\vec{\theta}}}$ and $\ket{\psi_{\vec{\phi}}}$. Subsequently, these states are simulated, and the fidelity, denoted as $F$, is computed as the squared overlap between them: $F = |\langle \psi_{\vec{\theta}}|\psi_{\vec{\phi}}\rangle |^2$~\cite{sim2019expressibility}. This entire process is iterated multiple times to generate multiple fidelity estimates from various parameter pairs. On the other hand, $P_{Haar}(F)$ is calculated using the analytical 
form of the probability density function of fidelities for the ensemble of Haar random states, $P_{Haar}(F) = (N-1){(1-F)}^{N-2}$ where $F$ is the fidelity, and $N$ is the dimension of the Hilbert space~\cite{zyczkowski2005average}. From Equation~\eqref{eq:expr}, we see that \emph{lower} KL divergence values indicate a closer approximation to the maximally expressive ensemble of Haar random states,~\ie, a \emph{more} expressible circuit. In other words, this implies a greater capacity to explore a wider range of states within the Hilbert space. The expected expressibility of a PQC for solving a specific problem depends on many factors, including the structure of the circuit, the choice of parameters, and the quantum state being represented.

% The most expensive and challenging part of expressibility computation is obtaining the estimated fidelity distribution from the target PQC. Given that this estimated fidelity relies on execution on a quantum backend, device noise becomes a substantial factor impacting the expressibility values of the PQC. Ongoing research is essential to address challenges and enhance the expressibility computations in the diverse quantum computing landscape. Recently, Aktar~\etal proposed an expressibility prediction model using machine learning techniques~\cite{aktar2023predicting}. 

The most challenging part of computing expressibility is obtaining the estimated fidelity distribution $P_{PQC}(F;\vec{\theta})$ from the target PQC. Sim~\etal~\cite{sim2019expressibility} used fidelity samples from 5,000 pairs of states for computing expressibilities in their experiment, which can be computationally expensive.
Moreover, since this estimated fidelity relies on execution on a quantum backend, device noise becomes another substantial factor impacting the expressibility value of the PQC. Ongoing research is essential to address challenges and improve the expressibility computations in the diverse quantum computing landscape. 
%Recently, an expressibility prediction model using machine learning techniques has been proposed~\cite{aktar2023predicting}.

\section{Proposed Framework}
\label{sec:framework}
In this section, we provide an overview of the proposed graph neural network-based framework for estimating the expressibility of PQCs shown in Figure~\ref{fig:diagram}. The framework consists of several key components, including dataset preparation, graph encoding, ground truth expressibility computation, and the training of a prediction model.
Leveraging these integral components, the proposed framework provides predictions for the expressibility of the PQC. The model and its capabilities significantly expand over an early prototype of ours~\cite{aktar2023predicting}.
%We first describe the dataset collection, which can be subdivided into three primary steps: random PQC generation, computation of expressibility, and the creation of a validation dataset.

\subsection{Dataset Collection}
\label{sec:dataset}
The dataset collection can be subdivided into three primary steps: random PQC generation, computation of expressibility, and creation of an evaluation dataset.

\subsubsection{Random PQC Generation} 
We create random parameterized quantum circuits using a diverse gate set, including single-qubit and multi-qubit gates. The circuit construction begins with an initial layer where random single-qubit gates (RX, RY, RZ, H, I) are applied to each qubit. Optionally, this layer may be followed by a second set of single-qubit gates. Subsequent layers introduce entanglement by randomly placing multi-qubit gates (CX, CRX, CRY, CRZ, SWAP) between randomly chosen pairs of qubits. The degree of entanglement, indicated by the number of consecutive multi-qubit gates, is also randomly determined. The algorithm further allows adding one or two-qubit layers comprising single-qubit gates. To deepen the circuit, we iterate through the layers of single-qubit gates and entanglements multiple times. The resulting dataset of random parameterized quantum circuits is highly adaptable and versatile, making it well-suited for a wide range of quantum computing applications. The random PQCs range from `1’ to `10’ qubits, repeated up to three times.

\subsubsection{Expressibility Calculation}
\label{sec:expr-calc}
We calculate the estimated fidelity distribution for each PQC in our dataset by randomly assigning two sets of parameters and then computing the squared overlap, representing quantum state fidelity between the resulting quantum states. First, each PQC is transpiled into native gates that refer to the target quantum device’s connectivity. To determine fidelity, we append the inverse circuit of the second state after the first circuit and subsequently measure the probability of obtaining the state $\ket{0}^{\otimes n}$, where $n$ represents the number of qubits~(Figure~\ref{fig:diagram}). We begin with the default initial state $\ket{0}^{\otimes n}$ and run the experiments on a quantum backend 5,000 times to attain a distribution of quantum state fidelities for each PQC. Then, we compute the distribution of Haar random state fidelities analytically (as described in Section~\ref{sec:expr}) using a histogram with a bin size of $75$ and calculate the expressibility value of the PQC using Equation~\eqref{eq:expr}. Since our prediction model considers the information from quantum circuits and noisy backends, we collect data from noiseless and noisy quantum backends. 

We generate 3,000 PQCs each on $1\leq n\leq8$ qubits, and 500 PQCs for $n=9,10$; for a total of \emph{25,000 random PQCs}. 
We compute each PQC’s expressibility using a noiseless IBM QASM simulator (to get the fidelity distribution). 
Furthermore, we collect the expressibility of \emph{4,000 PQCs on three noisy backends} (IBM FakeGuadalupe, FakeMumbai, and FakeHanoi) to investigate the impact of noise on expressibility prediction.  

\subsubsection{Evaluation Dataset}
For validation, we incorporate hardware-efficient ansatz sets from IBM Qiskit RealAmplitude circuits~\cite{realAmplitude}, tailored explicitly for systems of up to four qubits, with a maximum of four circuit repetitions. We adopt various entanglement patterns, such as full, linear, circular, and shifted-circular-alternating, and generate a comprehensive set of 64 hardware-efficient circuits. Furthermore, our datasets encompass 19 reference circuits, originally employed in prior experiments~\cite{sim2019expressibility}, and these are replicated with up to three circuit repetitions. We collect the evaluation dataset samples utilizing noiseless and noisy quantum backends.

\begin{figure*}[t!]
    \centering 
    \includegraphics[width=0.255\linewidth]{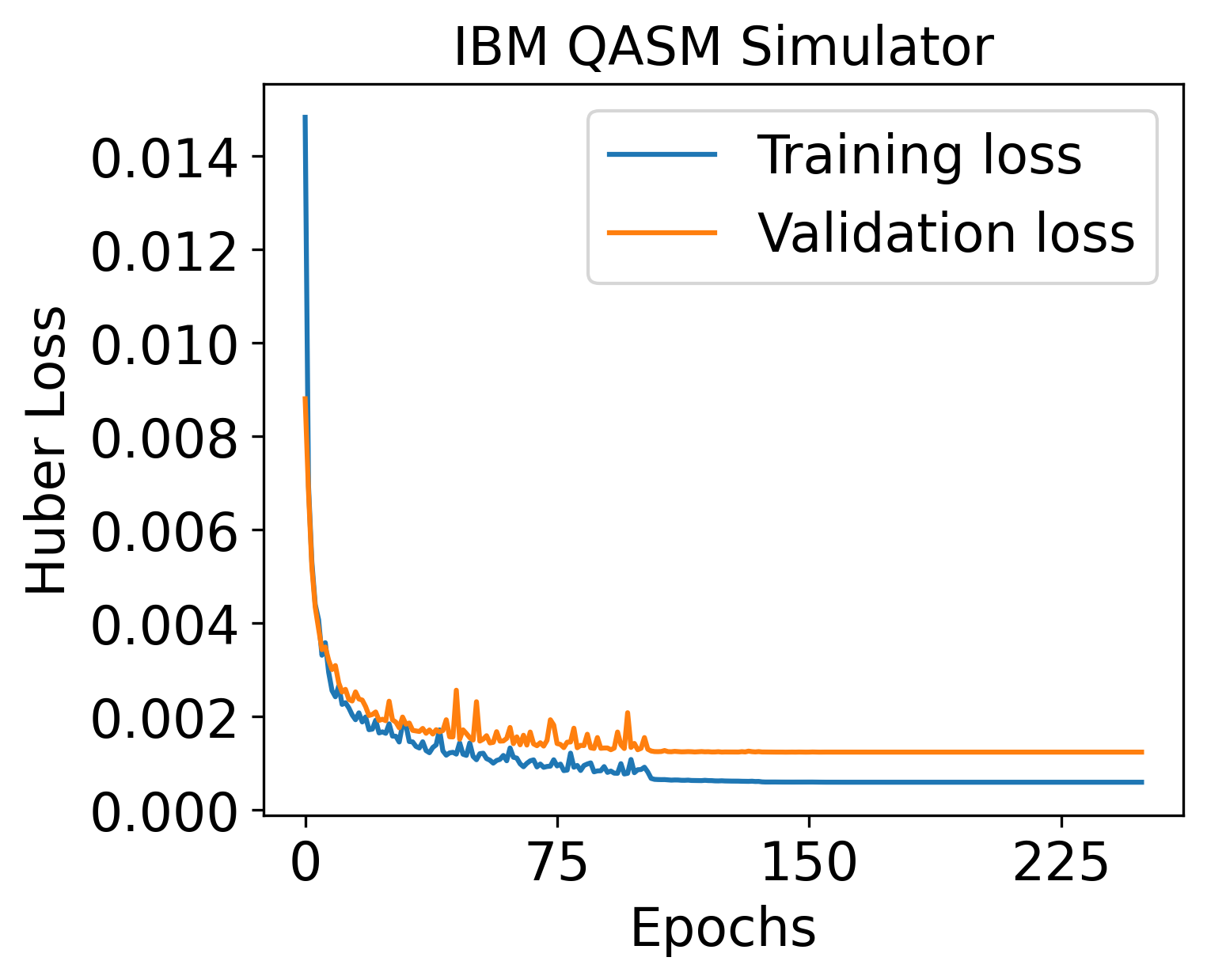}\hfill%
    \includegraphics[width=0.245\linewidth]{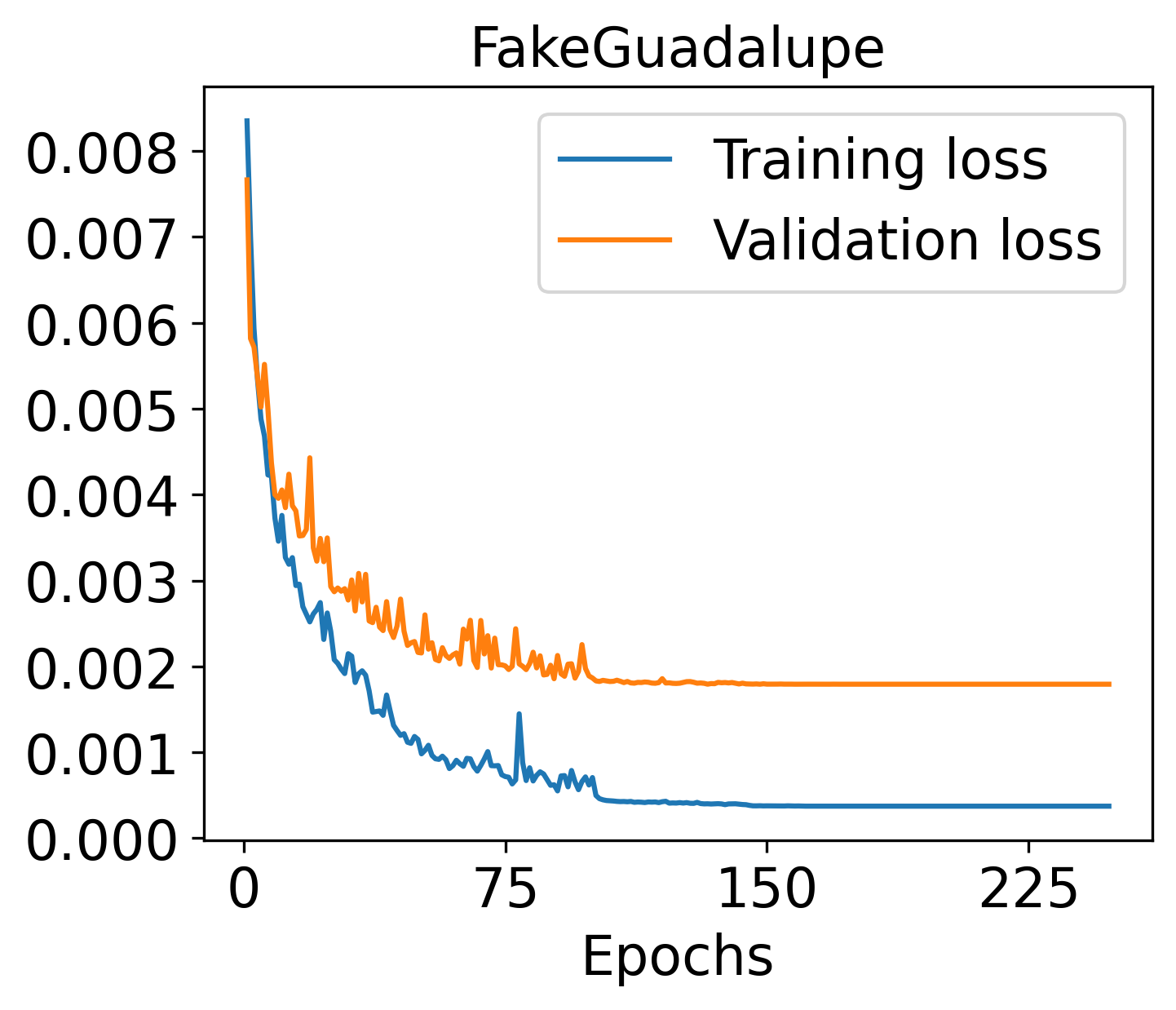}\hfill%
    \includegraphics[width=0.245\linewidth]{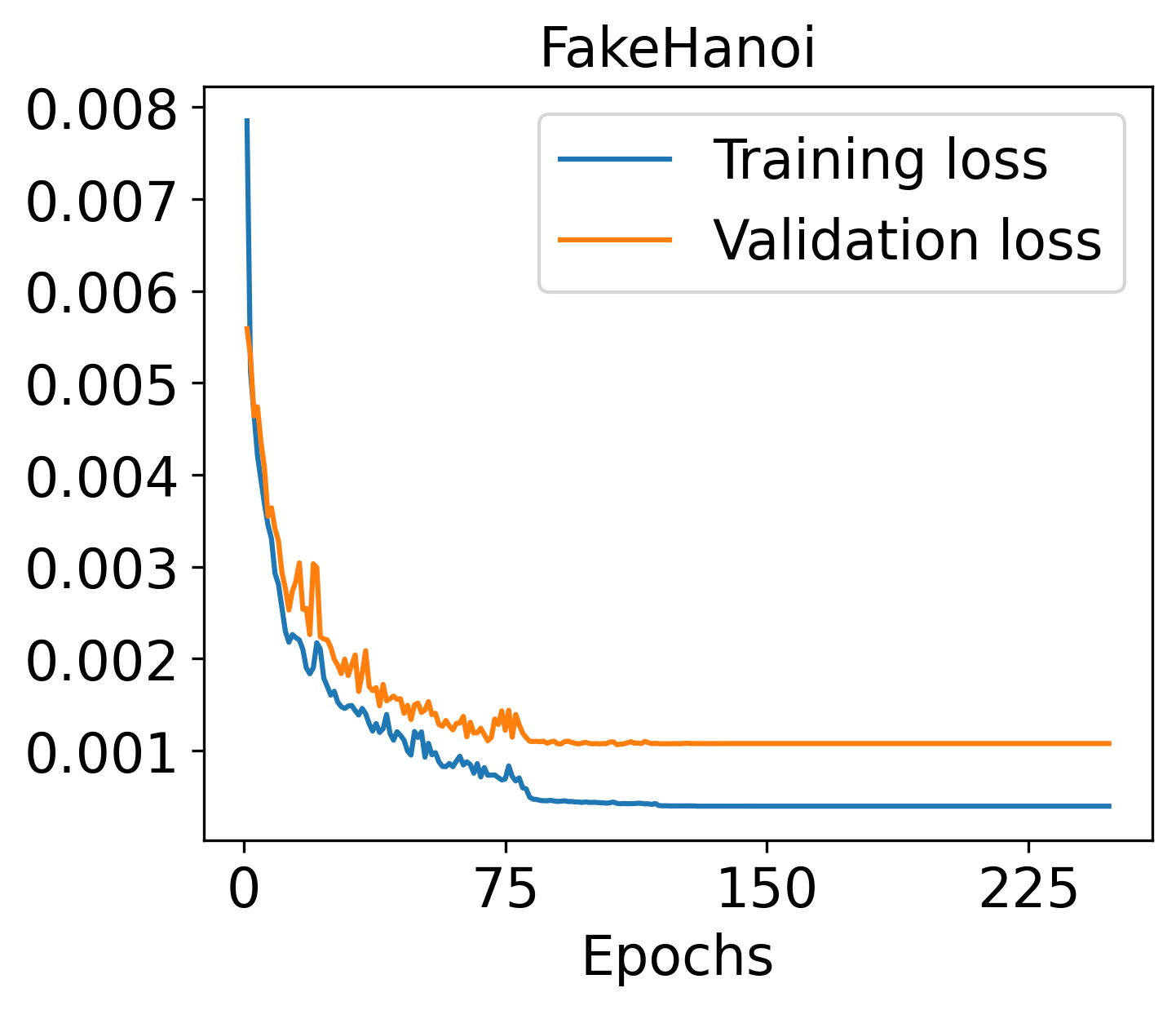}\hfill%
    \includegraphics[width=0.245\linewidth]{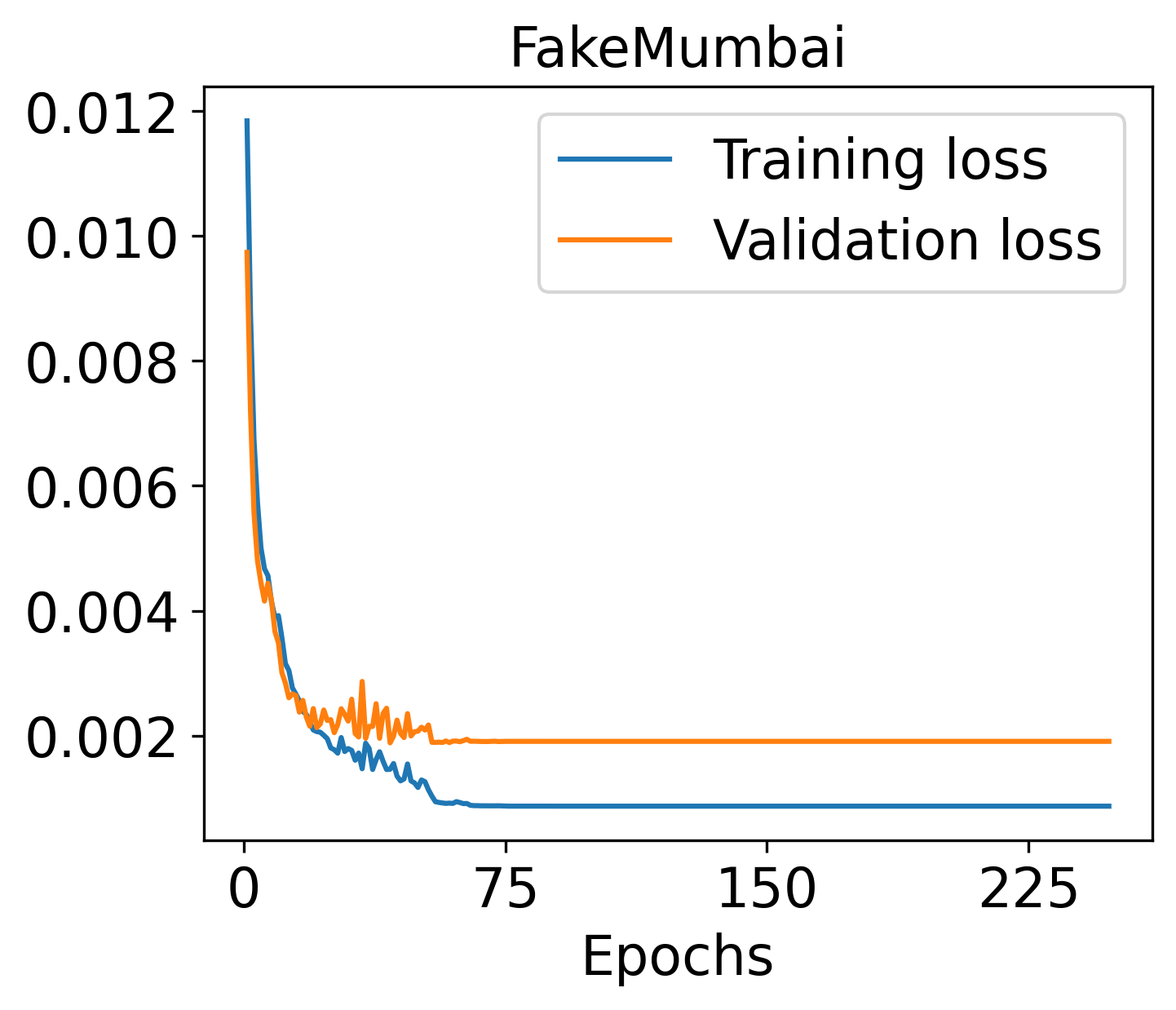}%
    \caption{Learning curves of our proposed GNN model on training and validation dataset of randomly generated PQCs on noiseless simulators and three noisy backends (FakeGuadalupe, FakeHanoi, and FakeMumbai). Each plot demonstrates consistent learning across different quantum backends.}
    \vspace*{1ex}
    \label{fig:training-result}
\end{figure*} 

\begin{figure*}[t!]
    \centering 
    \includegraphics[width=0.255\linewidth]{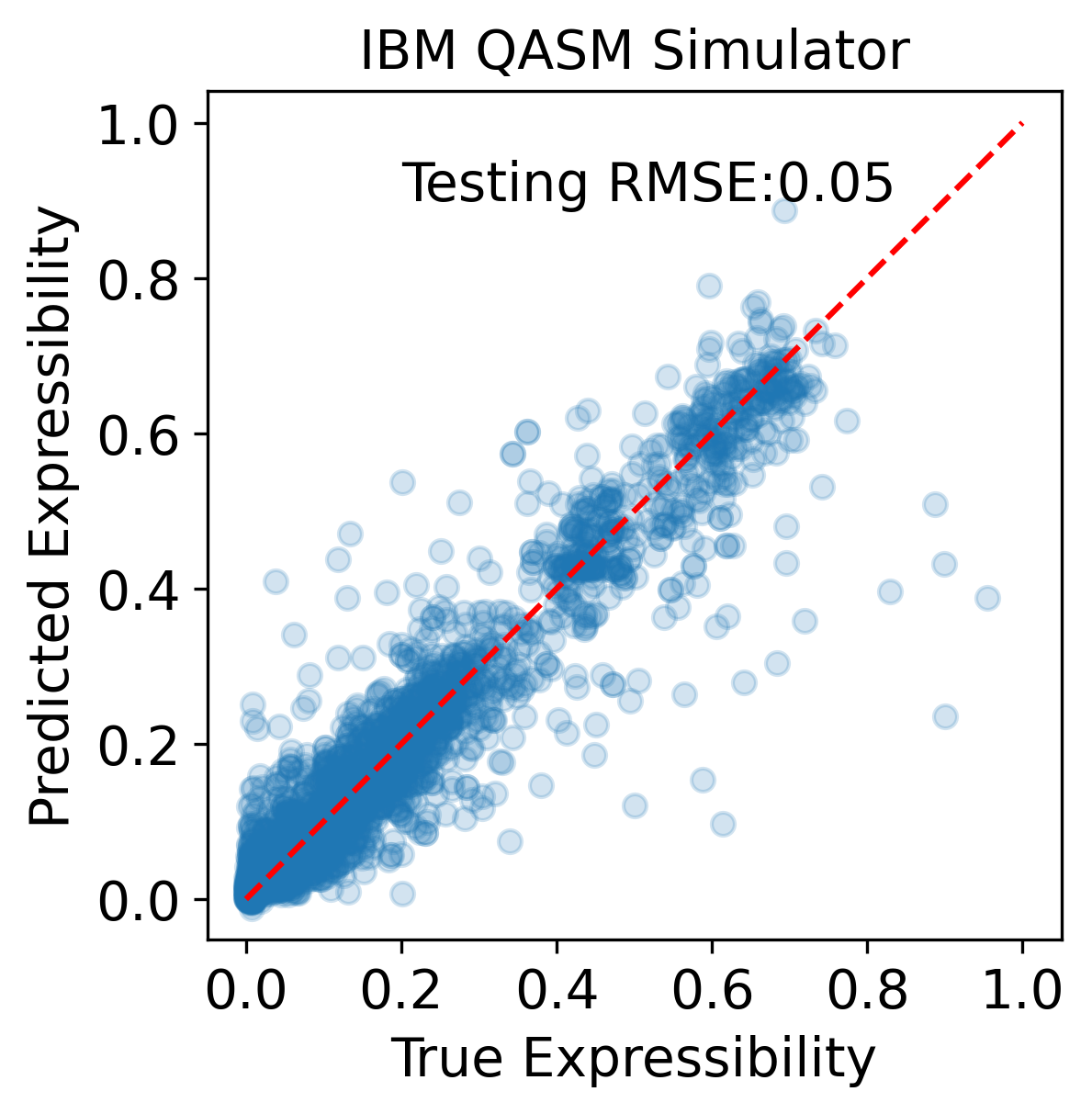}\hfill%
    \includegraphics[width=0.245\linewidth]{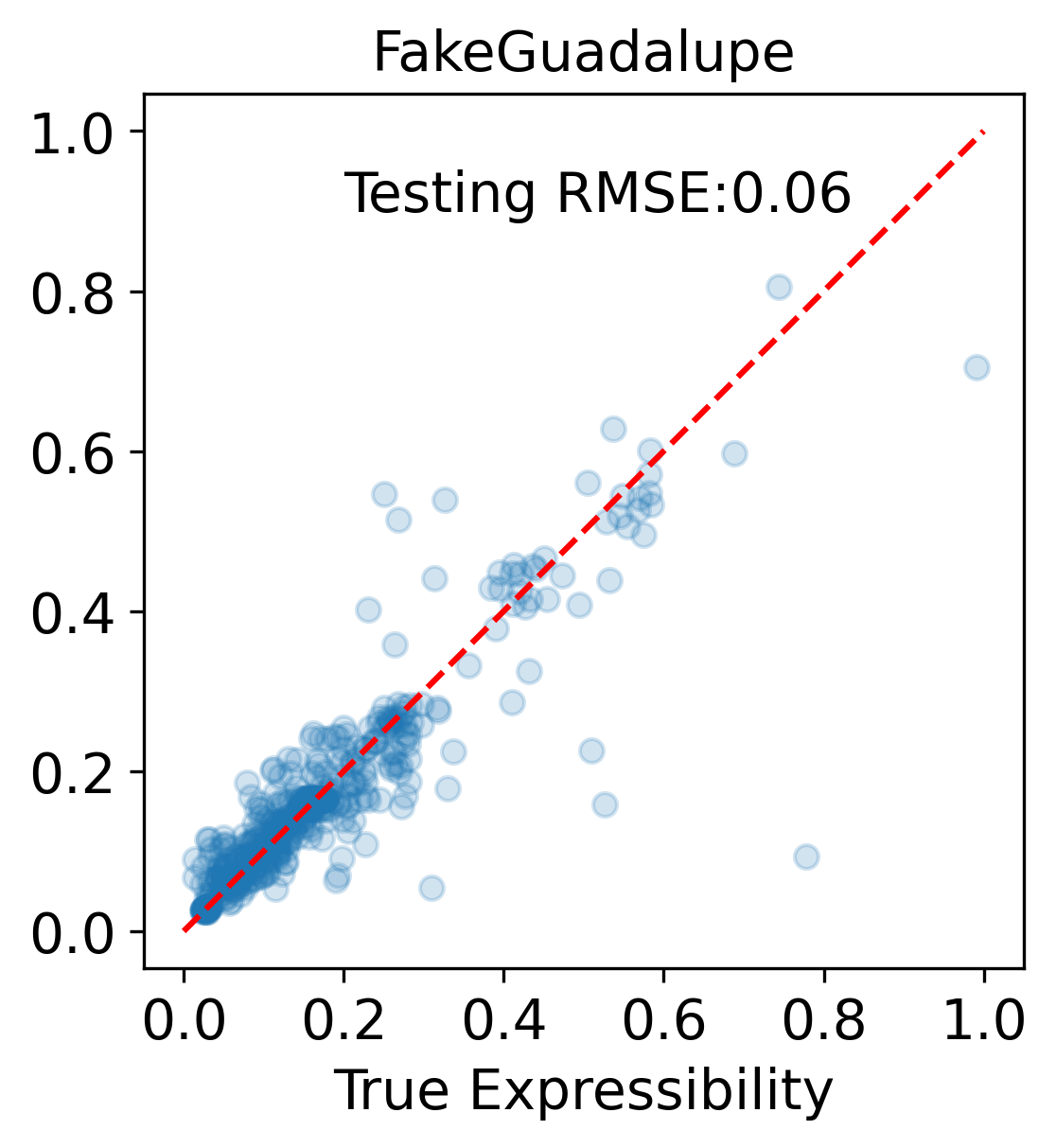}\hfill%
    \includegraphics[width=0.245\linewidth]{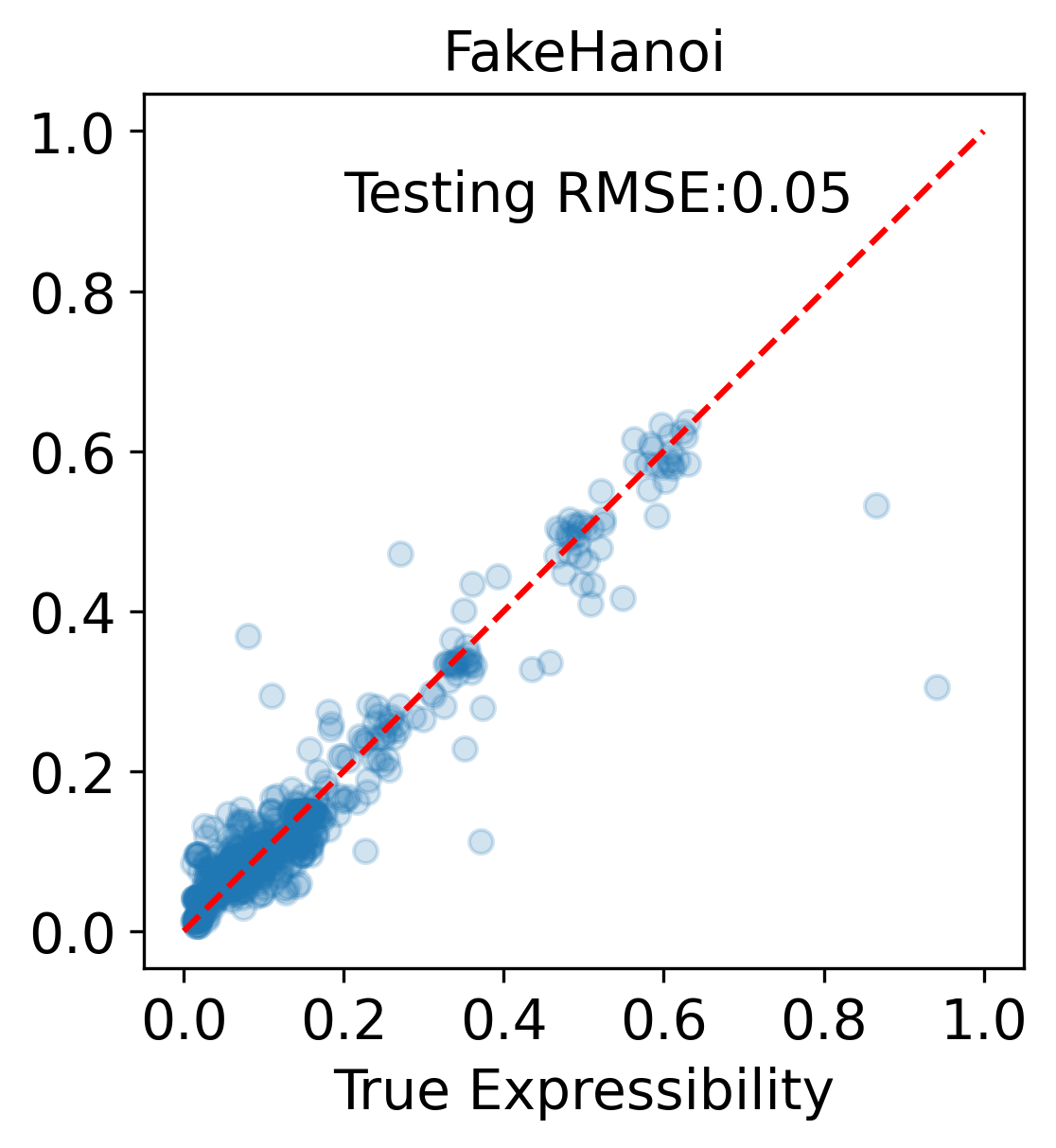}\hfill%
    \includegraphics[width=0.245\linewidth]{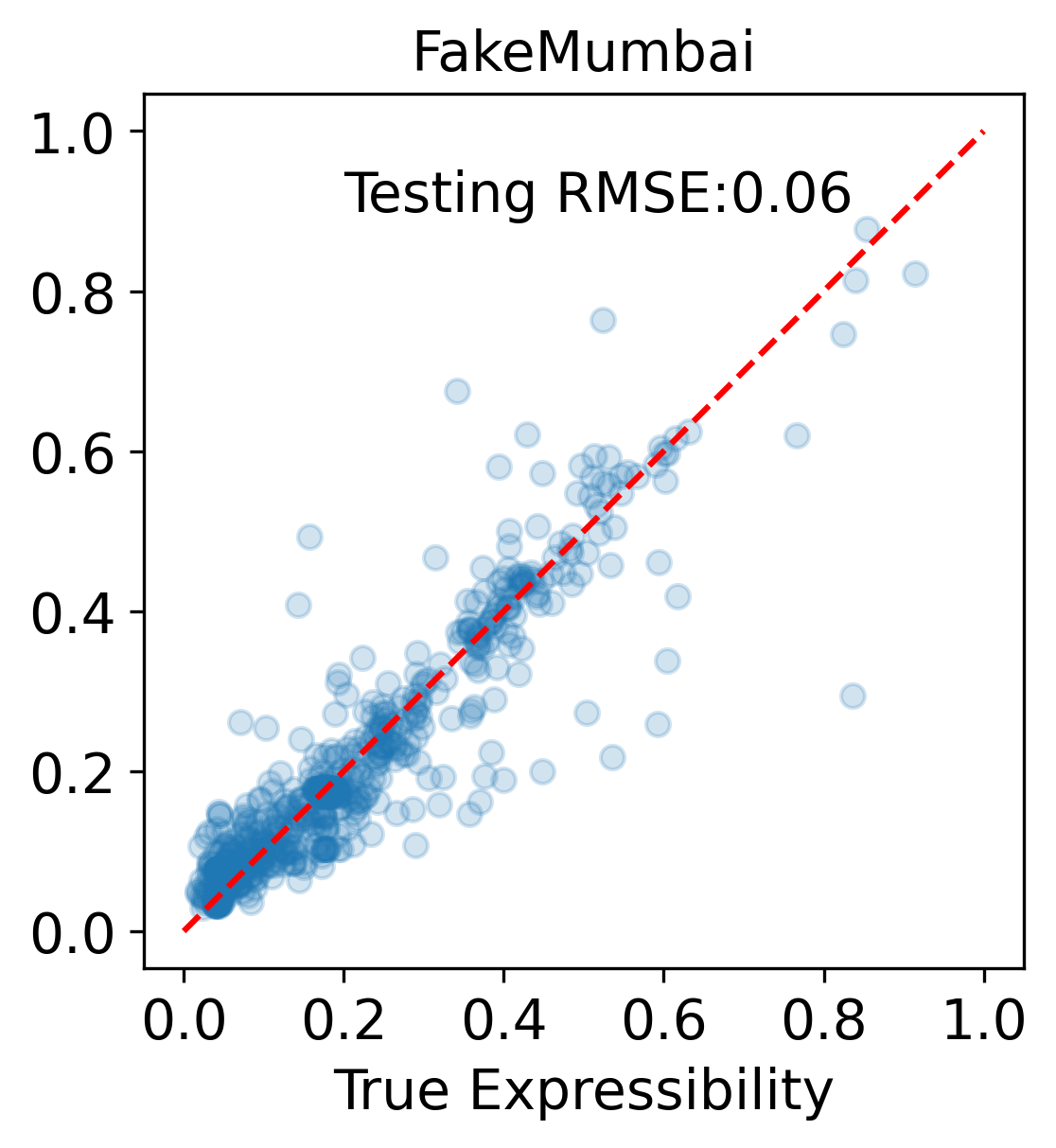}%
    \caption{Scatter plots of expressibility for randomly generated PQCs on noiseless simulators and three noisy backends (FakeGuadalupe, FakeHanoi, and FakeMumbai). Our GNN-based prediction model can provide close to accurate estimations of expressibility with RMSE $0.05$ in a noiseless IBM QASM simulator and RMSE lower than $0.06$ in the three noisy backends.}
    \label{fig:testing-result}
\end{figure*}

\subsection{Prediction Model}
We utilize the dataset described in the previous section to learn the complex relationship between PQCs and their expressibility. To this end, we extract features from the graph representation of the PQCs and employ them in the GNN model along with ground truth expressibility values. The various components of the GNN model capture the complex patterns and dependencies within the graph representation, facilitating the learning of this relationship. %between PQCs and their expressibility. 

\vspace*{1ex}
\subsubsection{Graph Extraction \& Encoding} 
We present each PQC as a directed graph $G = (V, E)$, where $V$ represents input nodes, gate nodes, and output nodes, while $E$ represents the set of edges indicating the flow of information within the PQC, as shown in Figure~\ref{fig:diagram}. Each node in the graph is associated with a feature vector encapsulating various features. These features include information such as the node type, the qubits utilized, and noise-related data, including parameters like T1 \& T2 for the associated qubits, gate errors, and readout errors, among others. The node type describes whether a node is an input, measurement, or a part of the device’s native gate set (RZ, X, SX, CNOT). The feature vector employed has a length of 23, with the first six digits corresponding to node types (one-hot encoded), the following ten digits representing the targeted qubit(s), and the last seven digits encoding backend noise information (see Figure~\ref{fig:node-feature}). In node types and targeted qubits, if a feature is irrelevant to a specific node, the corresponding value is assigned 0; otherwise, it is set to 1. Each PQC is represented as a list of dictionaries, with each dictionary corresponding to a feature vector and the number of dictionaries dependent on the nodes of the circuit. The circuits also include global feature sets,~\ie, circuit depth, number of parameterized gates, number of qubits, and counts of different gates represented as a vector.
Additionally, each PQC includes an edge matrix that captures the relationships between nodes in the graph. The computed expressibility value for each PQC is included as the target variable in the dataset. The machine learning model utilizes these to learn the relationship between the input features and the expressibility of the PQCs. 

\begin{figure*}[t!]
    \centering 
    \includegraphics[width=0.49\linewidth]{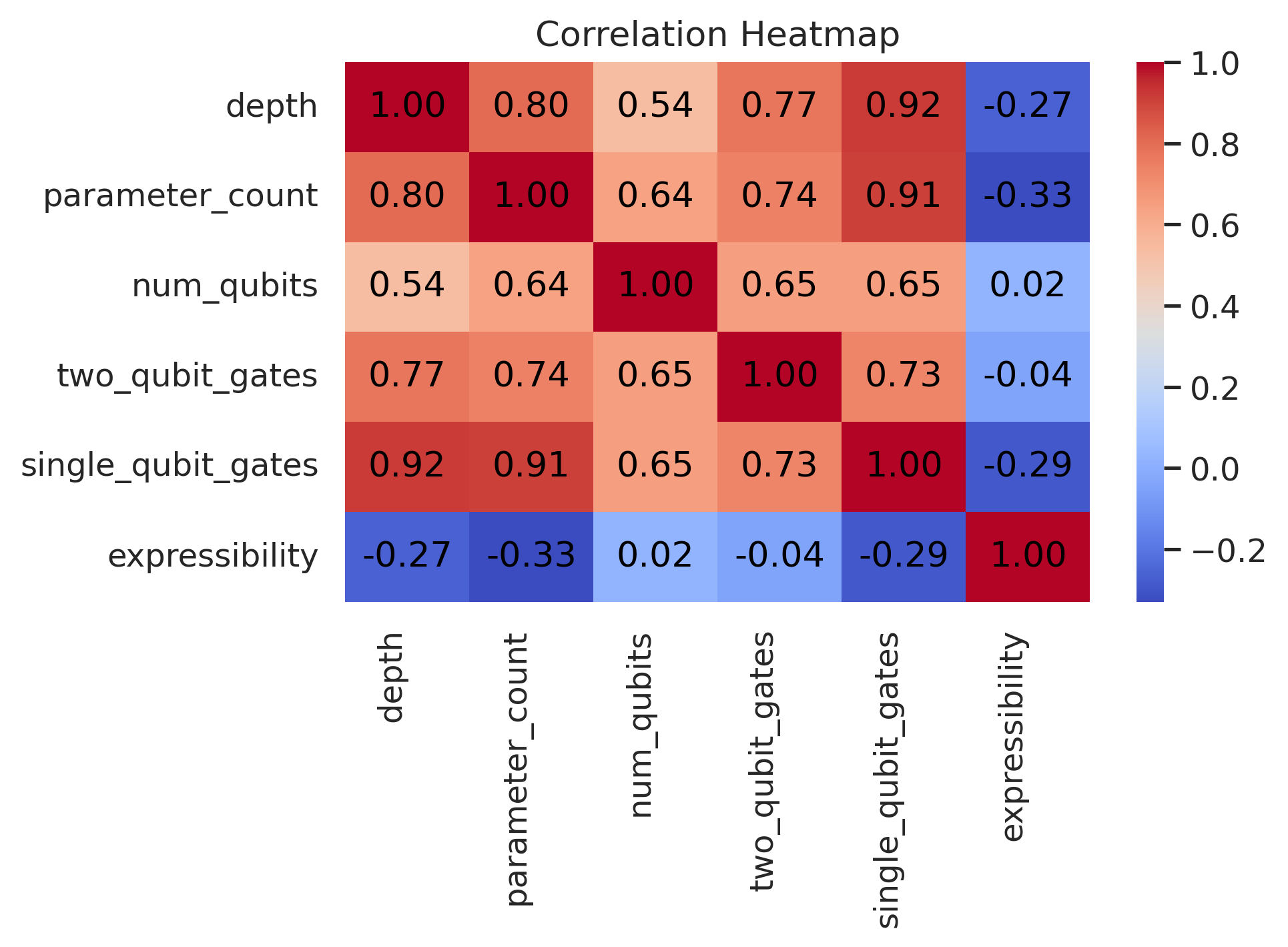}\hfill%
    \includegraphics[width=0.49\linewidth]{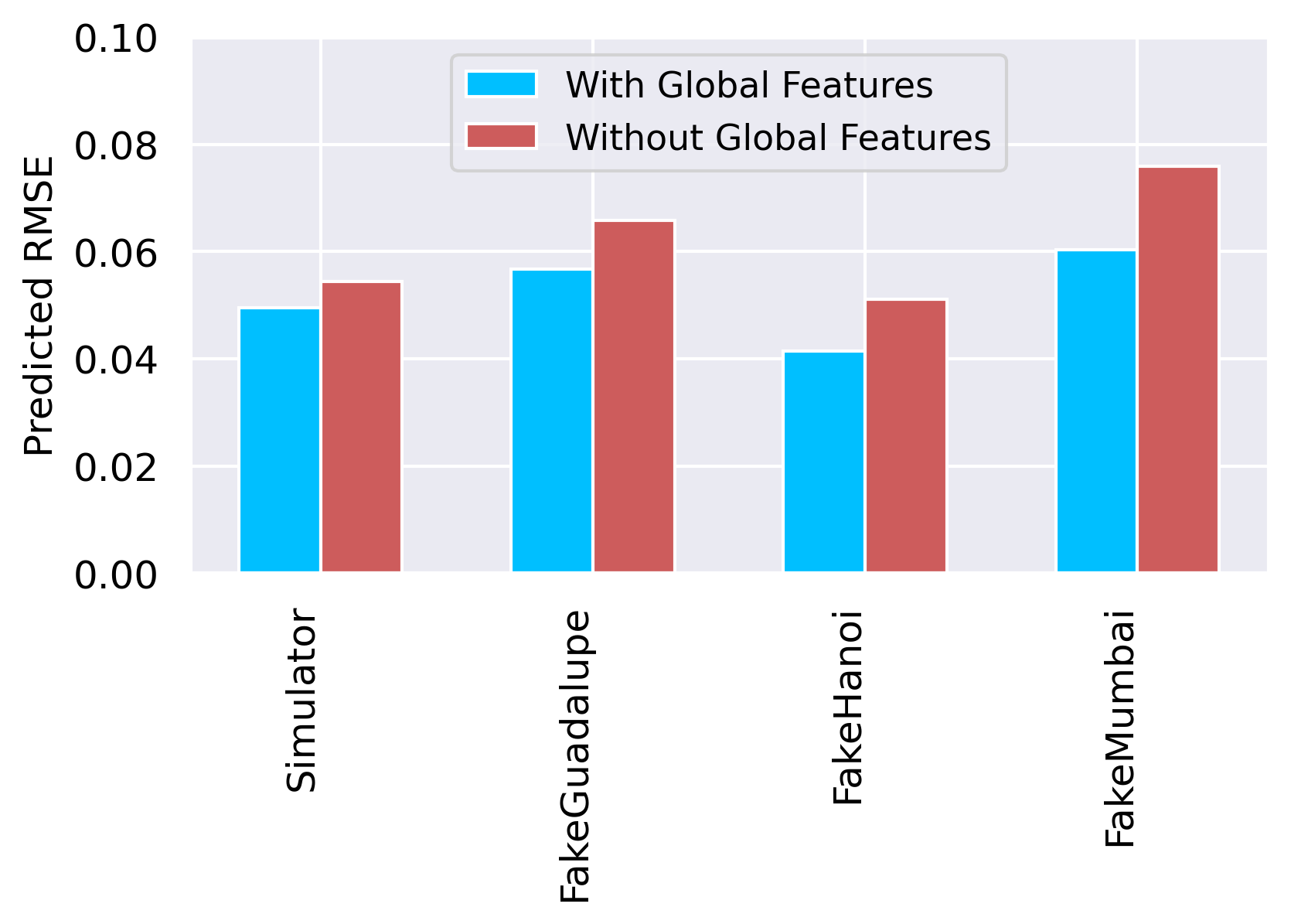}%
    \caption{\textit{(Left)} Heatmap displaying the correlation between global feature variables and the expressibility of PQCs. Notably, circuit depth, number of parameters, and number of single-qubit gates exhibit a good negative correlation with the expressibility of PQC. \textit{(Right)} Bar chart demonstrating the impact of global features on the RMSE on the testing dataset. Incorporating global features improves model accuracy (reduced RMSE loss) across all quantum backends. }
    \label{fig:input-feature-analysis}
\end{figure*}

\subsubsection{Graph Neural Network Model}
We employ a Graph Neural Network (GNN) model to learn and predict the relationship between PQC and expressibility. The model is adapted from the graph transformer model proposed by Wang~\etal~\cite{Wang2022torchQuantum} for estimating the reliability of quantum circuits. The overview of the GNN model is shown in Figure~\ref{fig:gnn}. The model starts with three TransformerConv~\cite{shi2020masked} layers to effectively capture and process the local neighborhood information within the PQC graphs. The initial layer is provided with the node features extracted from the graph representation of a PQC. 
Additionally, ReLU activation layers are incorporated between each of the TransformerConv layers to prevent potential numerical errors caused by negative values. The learned features are aggregated using \textit{Global Mean Pooling} to obtain an aggregated node feature vector. Besides, the global features of the circuits are propagated through three fully connected (FC) layers, with ReLU activation layers interspersed between them. Next, the global feature vector is concatenated with the aggregated node feature vector and fed to the \textit{regressor} consisting of three fully connected layers. The output is a single value corresponding to the predicted expressibility value of the PQC. During the backpropagation process, the model parameters are optimized to minimize the Huber loss between the predicted and the ideal expressibility values, ensuring the capability of the model to predict expressibility efficiently.

\section{Experimental Results}
\label{sec:exp}
\subsection{Experimental Setup}
\label{sec:exp-setup}
Our noiseless simulator dataset contains expressibility values for $25,000$ random PQC samples, and we also collect $4,000$ random PQC samples from each noisy backend (IBM FakeGuadalupe, FakeMumbai, and FakeHanoi). 
We partition the dataset of $8\times 3,000$ random PQCs on $n\leq 8$ qubits into three distinct subsets: $70$\% for training, $10$\% for validation, and $20$\% for testing. 
Each subset maintains an equal number of samples of PQCs on $n=1,\ldots,8$ qubits to ensure balance during the learning phase. 
This division allows us to perform training, validation, and testing on independent and representative portions of the data. 
The remaining $2\times 500$ PQC samples on $n=9,10$ qubits are used to test the extrapolation of our model to out-of-range circuits (Subsection~\ref{sec:extrapolation}).

Initially, we normalize the node features across the dataset by centering them around the mean and scaling by the standard deviation. Our training procedure utilizes the \textit{Adam} optimizer for $300$ epochs, with a learning rate of $10^{-4}$ and a weight decay of $10^{-6}$. We incorporate a \textit{ReduceLROnPlateau} scheduler that dynamically reduces the learning rate by a factor of $0.1$ to fine-tune the training process and enhance model convergence. We utilize a batch size of $1,500$ and employ the \textit{Huber loss} as our chosen loss function during the training process. We implement a model checkpointing mechanism to ensure optimal performance by saving the model with the best validation loss.

\begin{figure*}[t!]
    \centering 
    \includegraphics[width=0.57\linewidth]{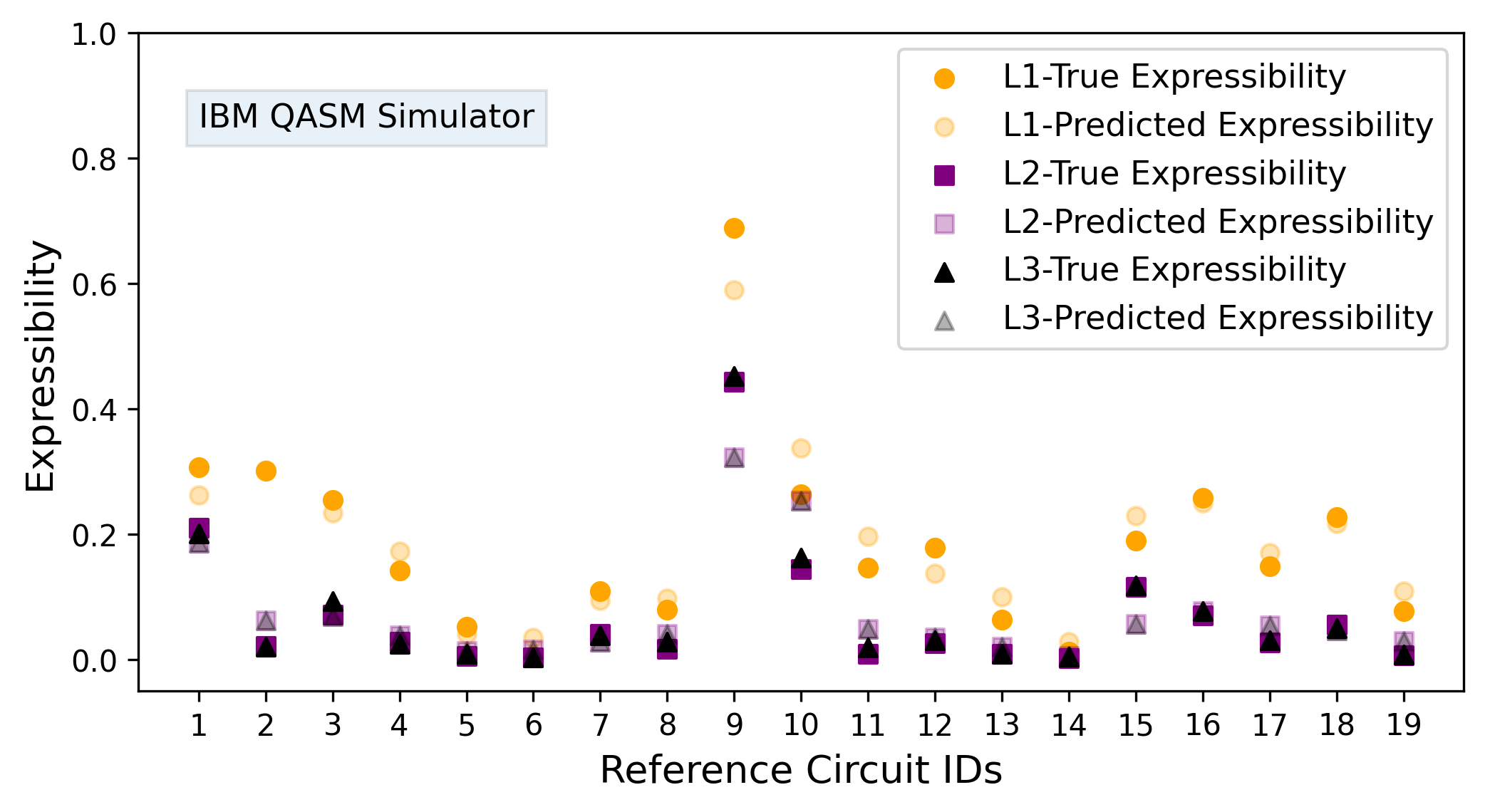}\hfill%
    \includegraphics[width=0.425\linewidth]{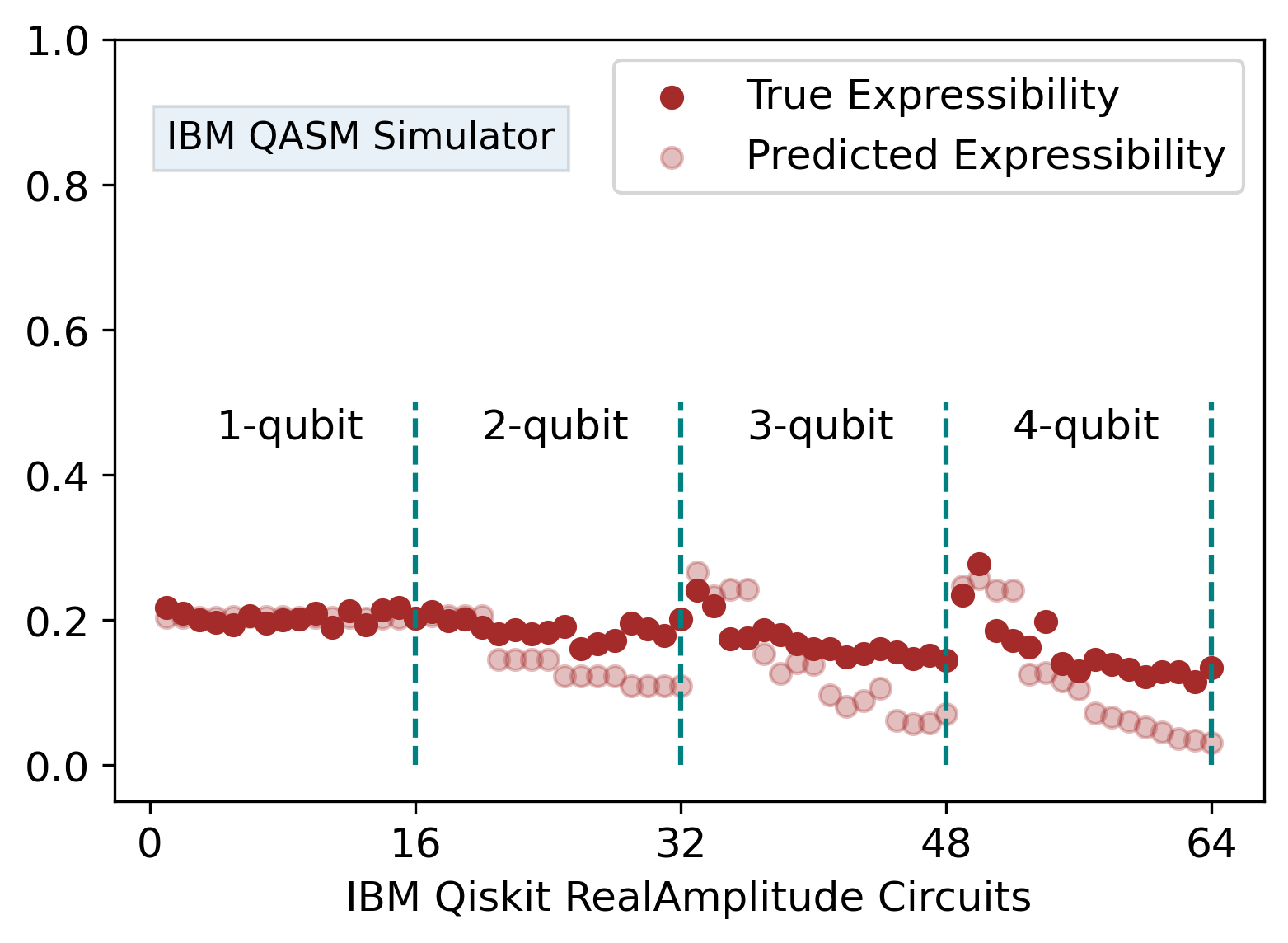}%
    \caption{Scatter plots of expressibility showing ground truth and predicted values in IBM QASM Simulator for 
    (\textit{left}) 19 reference circuits from Sim \etal~\cite{sim2019expressibility} with different number of base circuit repetitions, denoted as layer-1, -2, and -3 (L1, L2, L3) and for 
    (\textit{right}) 64 RealAmplitude circuit sets from IBM Qiskit~\cite{realAmplitude}. Both plots show close to true value expressibility prediction for PQCs achieving RMSE of $0.05$ and $0.06$, respectively.}
    \label{fig:sim-validation-result}
\end{figure*}

\subsection{Model Training}
We train the GNN model separately on the PQC samples collected from a noiseless IBM QASM simulator and noisy backends using the setup described in Section~\ref{sec:exp-setup}. Figure~\ref{fig:training-result} demonstrates learning of the model on the training and validation datasets of randomly generated PQCs. The first plot displays training and validation curves for the simulator dataset, while the subsequent plots depict the curves for three noisy backends. Each plot shows consistent learning patterns. The simulator training begins with lower loss compared to the noisy backends, but both converge similarly over epochs. 

After training, we evaluate the model performance on unseen testing samples. Figure~\ref{fig:testing-result} shows scatter plots of predicted expressibility vs true expressibility of randomly generated PQCs in the testing dataset. For each plot, the $X$-axis represents ground truth expressibility values while the $Y$-axis shows predicted expressibility values. The $y=x$ regression line represents the alignment between predicted and actual expressibility values, indicating how closely the predictions match the ground truth values. 

The leftmost plot illustrates predictions for the noiseless simulator where the predicted expressibility values closely match the actual values, resulting in an RMSE of $0.05$. The subsequent three plots show prediction for the FakeGuadalupe, FakeMumbai, and FakeHanoi backends, achieving RMSE values of $0.06$, $0.05$, and $0.06$, respectively. These plots show that the noiseless simulator offers better predictions while the model maintains decent performance even on noisy backends.

\begin{figure}[bh!]
    \centering
    \vspace*{-2ex}
    \includegraphics[width=0.7\linewidth]{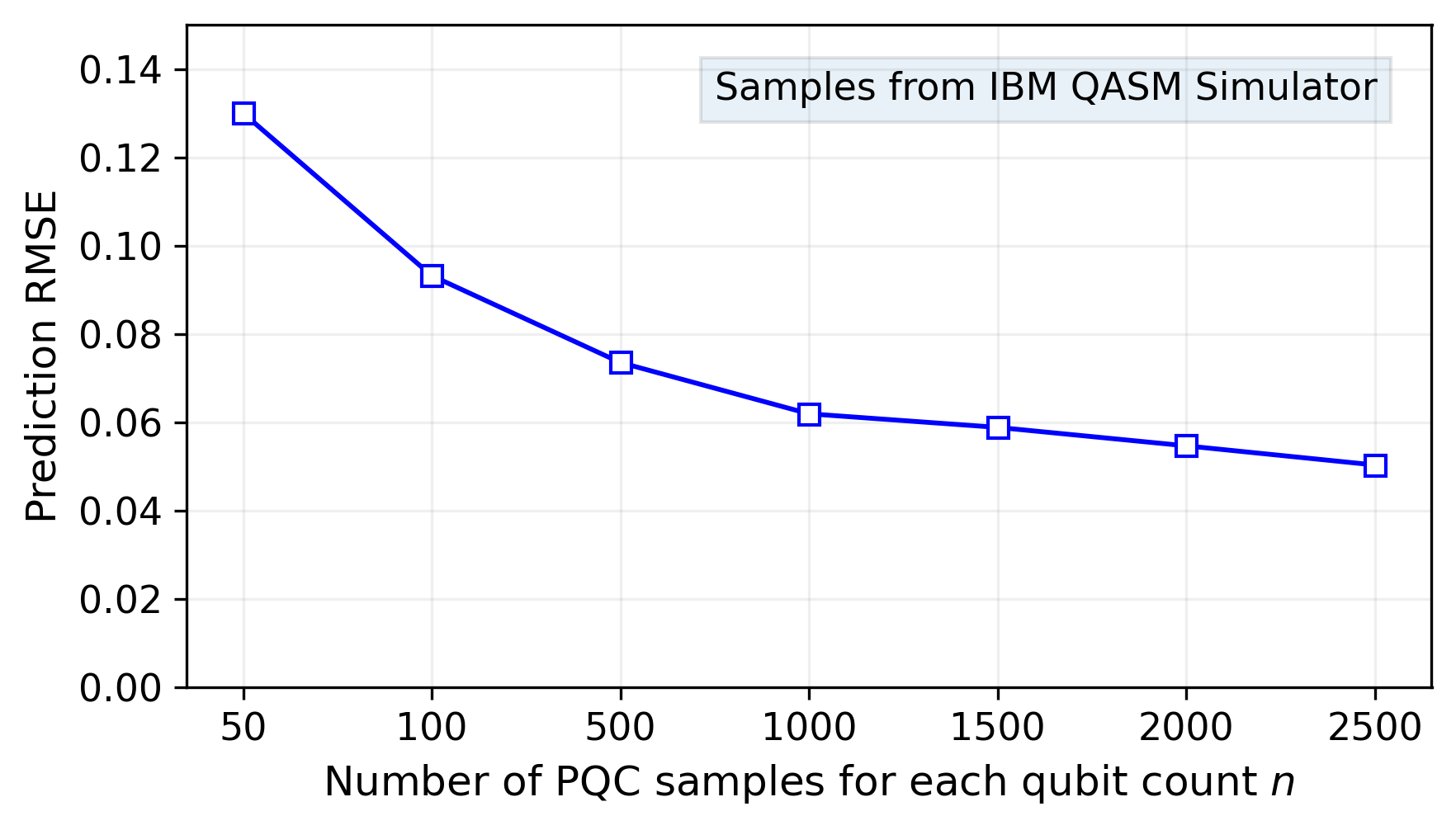}\\[-1ex]
    \caption{Demonstrating the effect of varied sample sizes on model learning performance using the prediction RMSE on the testing dataset of $8\times 500$ PQC samples. Learning PQC samples were chosen from noisy simulator datasets ranging from $8\times50$ to $8\times 2,500$ samples, equally balanced over circuits of size $1\leq n\leq 8$. The plot reveals a trend of decreasing prediction RMSE as the number of samples increases.%
    }
    \label{fig:sample-vs-rmse}
\end{figure}
\subsection{Model Learning Analysis}
We first investigate the correlation of the global circuit features with the target expressibility of a PQC. First, we compute a correlation heatmap that shows the pairwise correlation of each of the variables used as global circuit features. The heatmap is shown in Figure~\ref{fig:input-feature-analysis} (left). We find that circuit depth, number of parameters, and number of single-qubit gates show an excellent negative correlation with the expressibility of PQC. In contrast, others have close to zero correlation. The negative correlation suggests that an increase in these variables tends to decrease expressibility values, with values closer to 0 indicating better expressibility. To assess the contribution of these features to model accuracy, we train the GNN model both with and without the inclusion of global circuit features. The results are illustrated in Figure~\ref{fig:input-feature-analysis} (right). The plot demonstrates that incorporating global circuit features alongside node features reduces the RMSE loss on the testing dataset for both noiseless and noisy backends. The model accuracy notably improves for the noisy backends.

Additionally, we analyze the impact of varying sample sizes on the model’s learning performance using samples from a noisy simulator. We uniformly select an equal number of samples from each qubit and gradually increase the sample size. The testing dataset comprises a fixed 4,000 samples, with 500 samples of PQCs on $1\leq n\leq 8$ qubits each. We train the model using a \emph{variable training sample size} (equally balanced over the eight different numbers of qubits $n$ used in the PQCs) and evaluate its performance on the testing set. Figure~\ref{fig:sample-vs-rmse} illustrates the relationship between sample size and model performance, demonstrating how the prediction RMSE decreases with increasing sample sizes. We find that with only $8\times500$ samples, the model achieves an RMSE of $0.073$. 
Furthermore, we observe a slight improvement in model accuracy as more samples are included. This underscores the effectiveness of our proposed expressibility prediction approach, as the model can provide reliable estimations with a limited number of samples per PQC size $n$. The analysis suggests that the model can be seamlessly extended to more qubits in a scalable manner in the number of samples.

\begin{figure*}[t!]
    \centering 
    \includegraphics[width=0.34\linewidth]{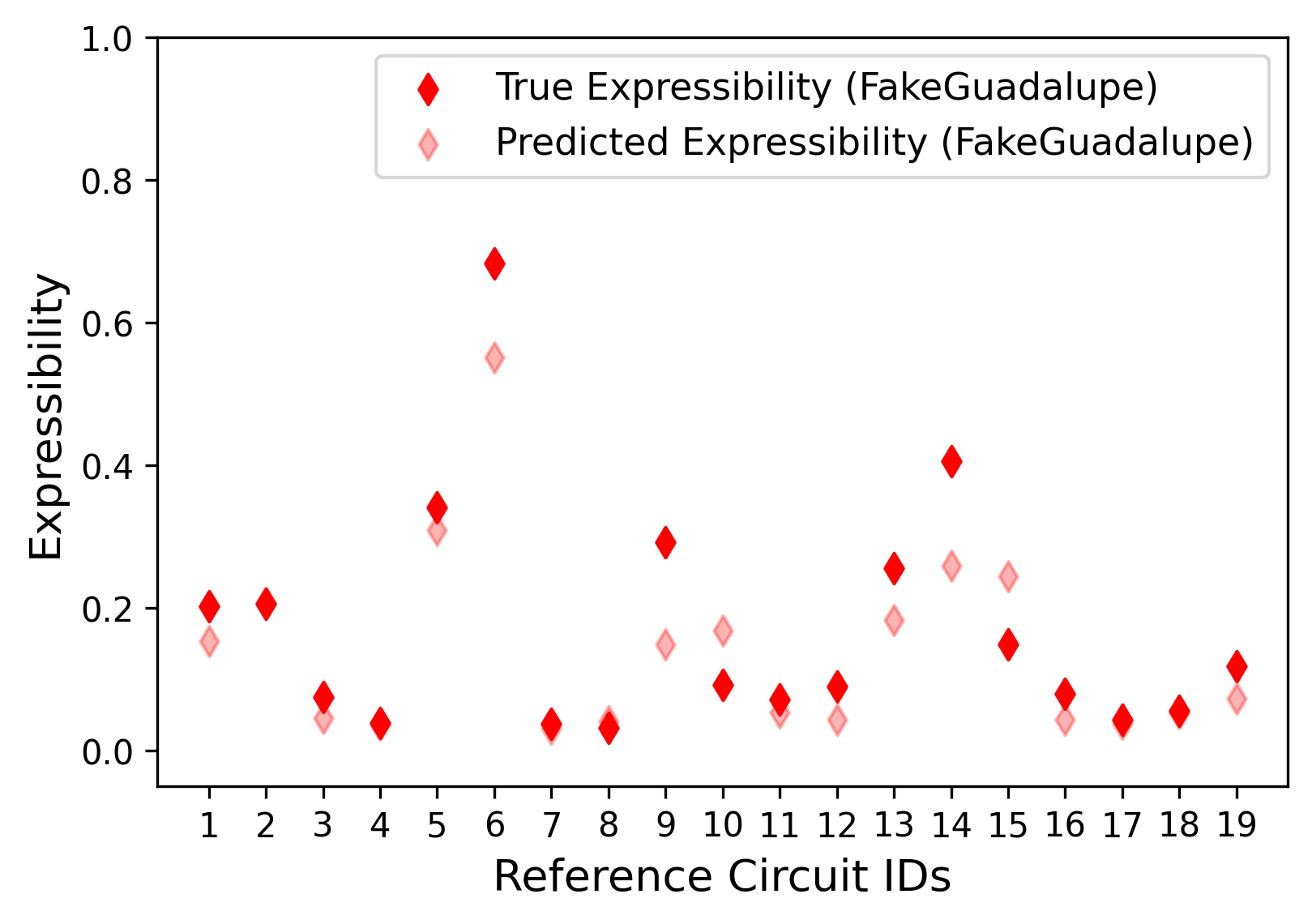}\hfill%
    \includegraphics[width=0.325\linewidth]{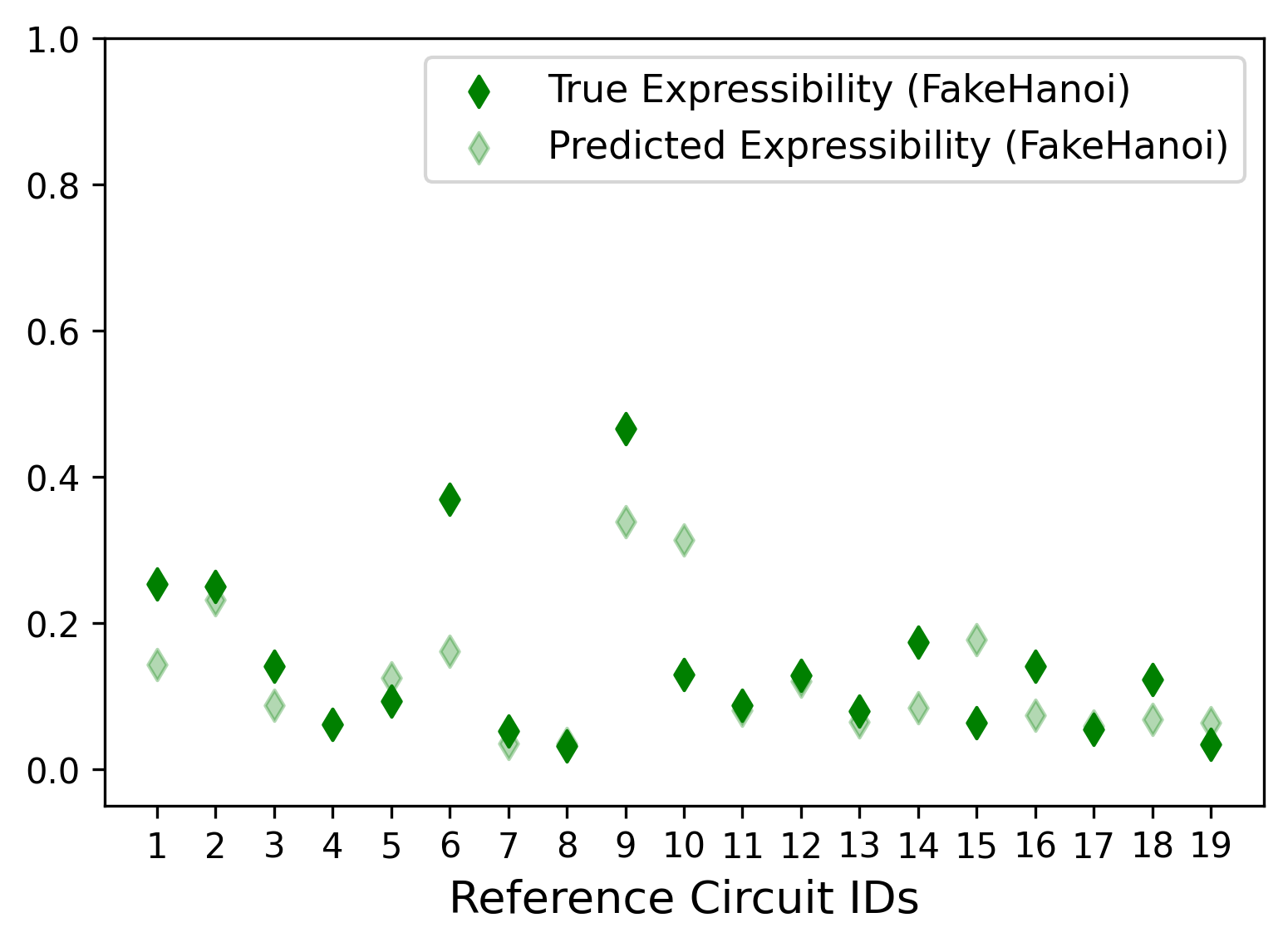}\hfill%
    \includegraphics[width=0.325\linewidth]{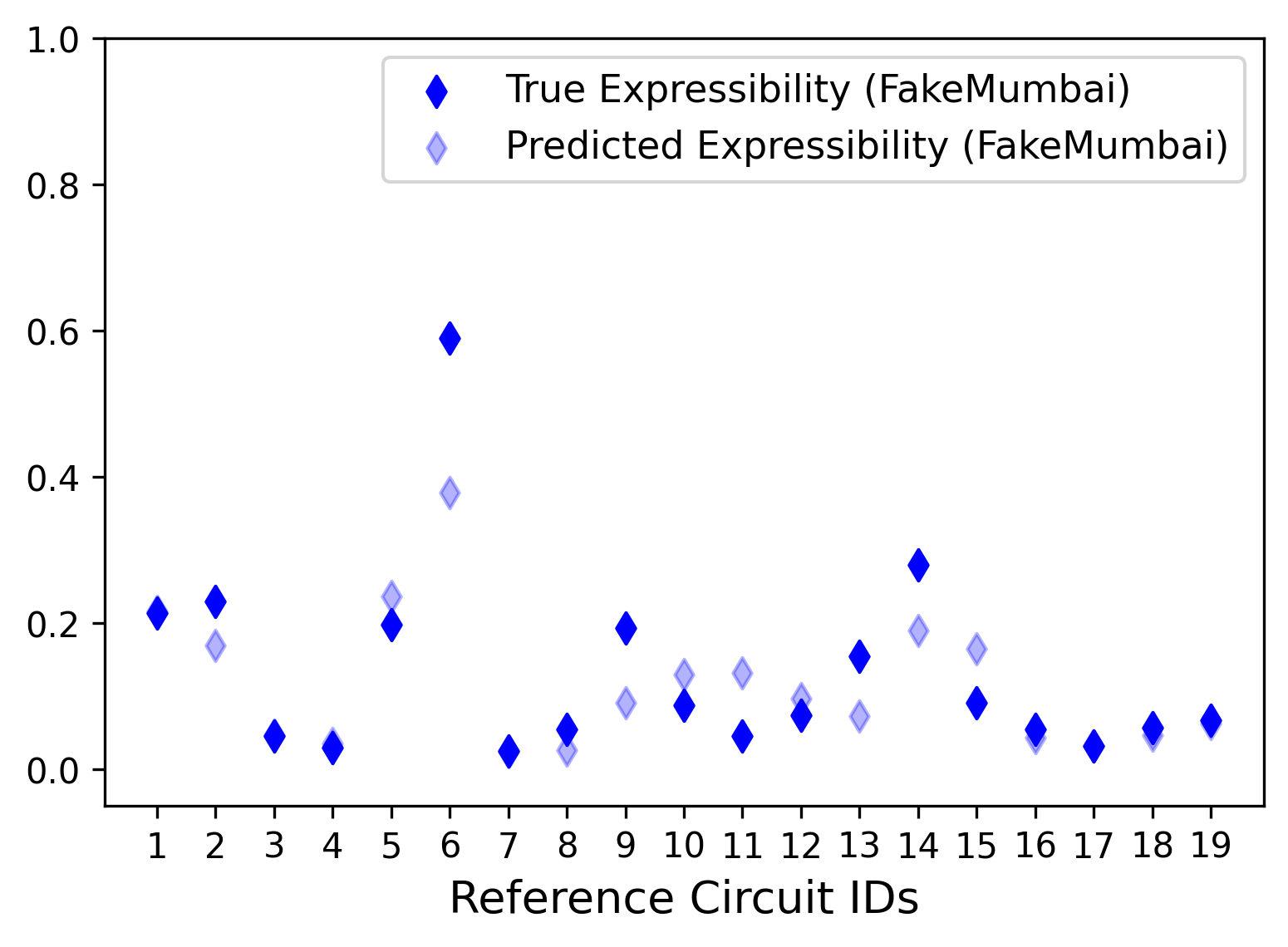}%
    \caption{Scatter plots of expressibility showing ground truth and predicted values of a single layer of original 19 reference circuits from~\cite{sim2019expressibility} in three noisy backends. The model shows good prediction in all three backends, achieving overall RMSE of $0.07$, $0.08$, and $0.07$ in FakeGuadalupe, FakeHanoi, and FakeMumbai backends, respectively.}
    \vspace*{1ex}
    \label{fig:validation-noise-result-ref19}
\end{figure*}  

\begin{figure*}[t!]
    \centering 
    \includegraphics[width=0.34\linewidth]{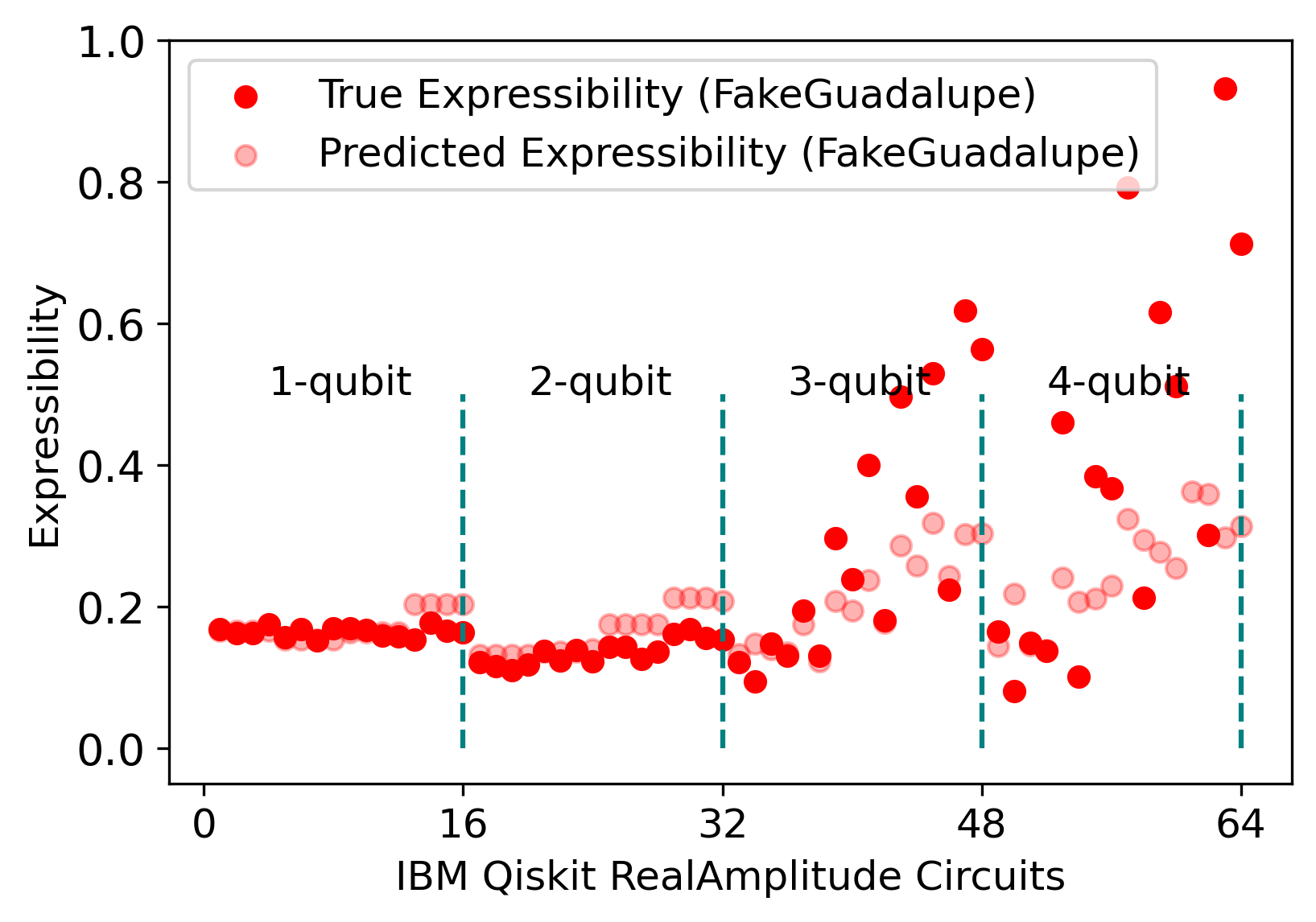}\hfill%
    \includegraphics[width=0.325\linewidth]{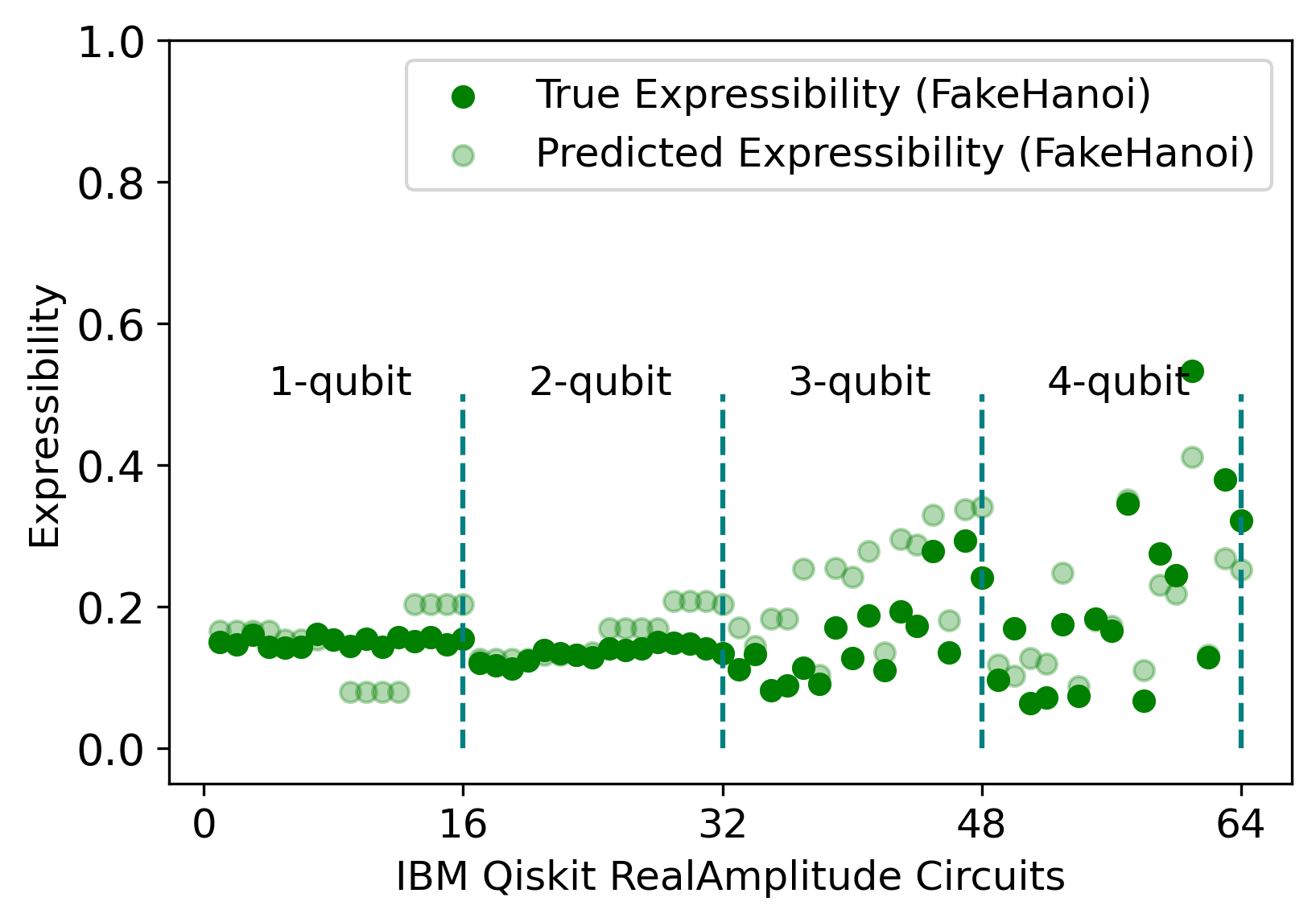}\hfill%
    \includegraphics[width=0.325\linewidth]{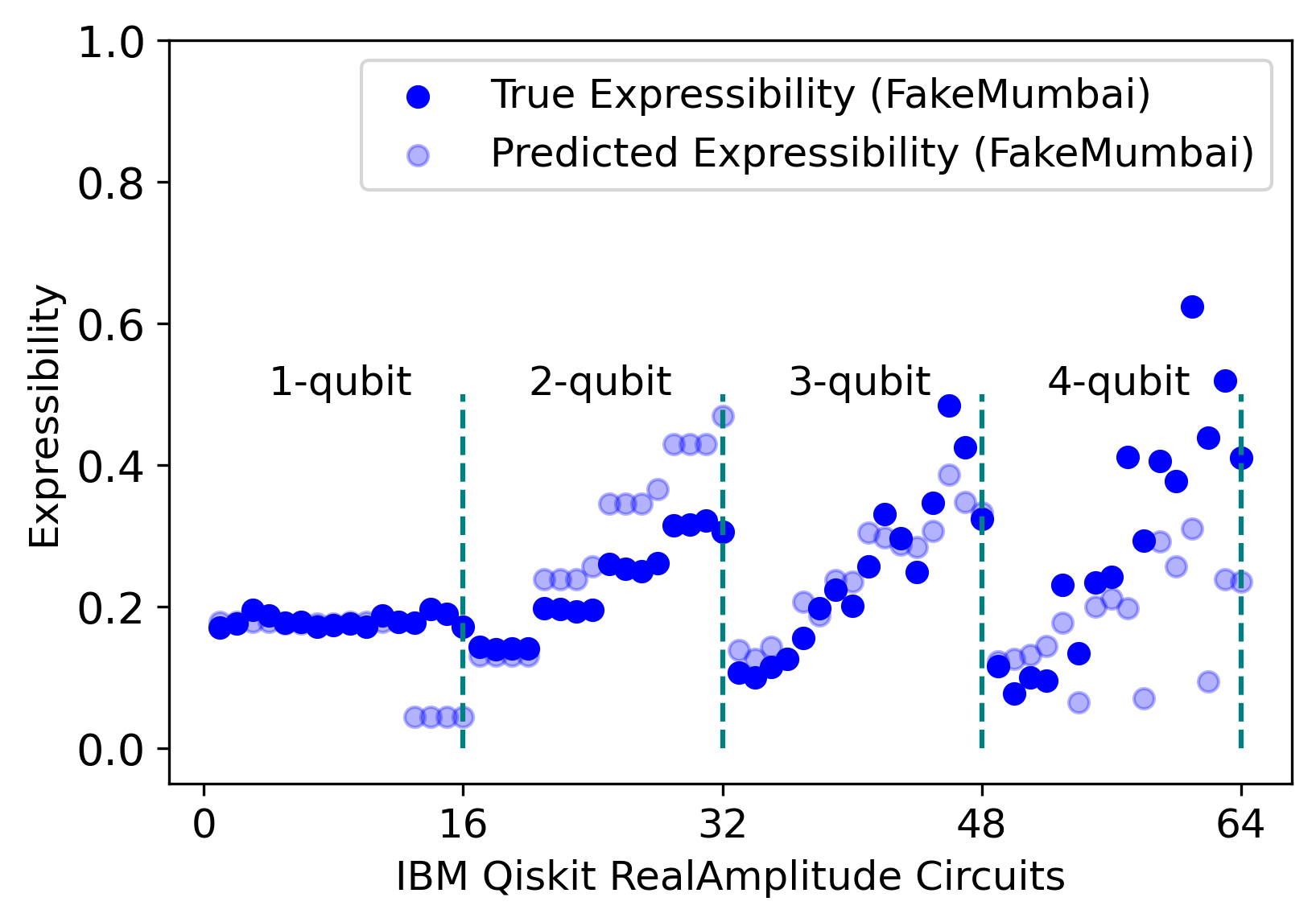}%
    \caption{Scatter plots of expressibility showing ground truth and predicted values of the 64 RealAmplitude circuit sets from IBM Qiskit in three noisy backends. The model shows good prediction in all three backends, achieving overall RMSE of $0.09$, $0.05$, and $0.09$ in FakeGuadalupe, FakeHanoi, and FakeMumbai backends, respectively.}
    \label{fig:validation-noise-result-realamp}
\end{figure*}

\subsection{Result Validation}
As there is no prior work on utilizing ML techniques to predict the expressibility of PQCs, we further validate the prediction model using two different sets of PQCs. We choose the 19 reference circuits from \cite{sim2019expressibility} and 64 PQCs from IBM Qiskit’s RealAmplitude circuit sets~\cite{realAmplitude} to compare the predictions of the trained model for both noiseless and noisy scenarios. Figure~\ref{fig:sim-validation-result} shows expressibility prediction for both validation sets. We represent ground truth expressibility values with solid color and predicted expressibility values with faded color. For each plot, the $X$-axis represents each circuit with IDs, and the $Y$-axis shows expressibility values for the PQCs. 
Figure~\ref{fig:sim-validation-result}~(left) shows expressibility prediction for the reference circuits from~\cite{sim2019expressibility} with different layers. In this context, \emph{layer} refers to the number of repetitions of the base circuit, denoted as layer-1 (L1), layer-2 (L2), and layer-3 (L3). In the left plot, \emph{orange} circles denote expressibility values for single-layer (L1) versions of the original 19 PQCs, \emph{purple} squares represent L2, and \emph{black} triangles represent L3 versions of the original 19 reference PQCs. From the plot, we see that adding multiple layers improves the expressibility of the PQCs, resulting in an increased capability to explore the Hilbert space. The plot shows that our predicted expressibility values are almost close to actual expressibility values for all the PQCs with different layers, achieving an RMSE of 0.05.
Figure~\ref{fig:sim-validation-result}~(right) shows expressibility prediction for the RealAmplitude circuit sets of IBM Qiskit. In this visualization, each set of 16 circuits, separated by vertical lines, corresponds to circuits with 1, 2, 3, or 4 qubits, respectively. We find that the predicted expressibility values are close to actual expressibility values but note that the 4-qubit circuits exhibit more disparity between predicted and actual values. Despite this, the model maintains reasonably accurate predictions across the different circuit configurations and achieves an overall RMSE of $0.06$.

%\begin{table}[t!]
%    \centering
%    \caption{Expressibility of five circuits~\cite{sim2019expressibility} on IBM-Hanoi with predictions from trained FakeHanoi and combined model.}
%    \vspace{-2mm}
%    \begin{tabular}{c|c|c|c|c|c}
%    \textbf{Expressibility} & \textbf{ID-3} & \textbf{ID-7} & \textbf{ID-9} & \textbf{ID-11} & \textbf{ID-15}\\
%    \hline   
%    \hline   
%    IBM-Hanoi & 0.280 & 0.237 & 0.451 &  0.266 & 0.302\\
%    \hline
%    Predicted (FakeHanoi) & 0.497 & 0.262 & 0.524 & 0.395 & 0.639 \\
%    \hline
%    Predicted (Combined) & 0.402 & 0.08 & 0.330 & 0.237 & 0.401\\
%    \end{tabular}
%    \label{tab:exp-compare}
%\end{table}
%\vspace{-4mm}

Figure~\ref{fig:validation-noise-result-ref19} shows the expressibility prediction of the original 19 reference circuits from~\cite{sim2019expressibility} in three noisy backends. The actual expressibility values of the $19$ reference circuits in a noiseless simulator and noisy backends are different. This shows the effect of noise in expressibility computation and the effect of solving variational quantum algorithms in noisy quantum backends. The plots show that our trained model can predict the expressibility values under different noise levels. The prediction achieved an overall RMSE of $0.07$, $0.08$, and $0.07$ in FakeGuadalupe, FakeHanoi, and FakeMumbai backend, respectively. 
Additionally, we calculate the actual expressibility values for reference circuits 3, 7, 9, 11, and 15 from the study by Sim et al. (\cite{sim2019expressibility}) on the IBM Hanoi hardware. Table~\ref{tab:exp-compare} compares the actual expressibility values obtained from the IBM Hanoi hardware and the predicted values generated by our trained model. We train two models: one using samples from all three noisy backends and another using samples from only the FakeHanoi backend. We find that the model trained with samples with different noise levels gives better prediction (overall RMSE $0.1$) than the model trained with only FakeHanoi samples (overall RMSE $0.19$). This indicates the potential of achieving good expressibility predictions on actual quantum backends, even when trained on samples from different fake backends.

\begin{table}[b!]
    \centering
    \vspace*{-1ex}
    \caption{Expressibility of five circuits from~\cite{sim2019expressibility} on IBM-Hanoi with predictions from trained FakeHanoi and combined model.}
    \begin{tabularx}{\linewidth}{@{\extracolsep{\fill}}lrrrrr@{}}
        \toprule
        \textbf{Expressibility} & \textbf{ID-3} & \textbf{ID-7} & \textbf{ID-9} & \textbf{ID-11} & \textbf{ID-15}\\
        \midrule
        IBM-Hanoi & 0.280 & 0.237 & 0.451 &  0.266 & 0.302\\
        Predicted (FakeHanoi) & 0.497 & 0.262 & 0.524 & 0.395 & 0.639 \\
        Predicted (Combined) & 0.402 & 0.080 & 0.330 & 0.237 & 0.401\\
        \bottomrule
    \end{tabularx}
    \label{tab:exp-compare}
\end{table}

Similarly, Figure~\ref{fig:validation-noise-result-realamp} demonstrates the expressibility prediction of RealAmplitude circuit sets from IBM Qiskit across three noisy backends. We observe a significant impact of hardware noise on the expressibility of these hardware-efficient ansatz sets. However, our trained models successfully predict the expressibility of these PQCs across all three noisy backends. Overall, we achieve an RMSE of $0.08$ for the FakeGuadalupe, FakeHanoi, and FakeMumbai backends.%, respectively.

\subsection{Extrapolation Analysis}
\label{sec:extrapolation}
\begin{figure}[t!]
    \centering 
    \includegraphics[width=0.7\linewidth]{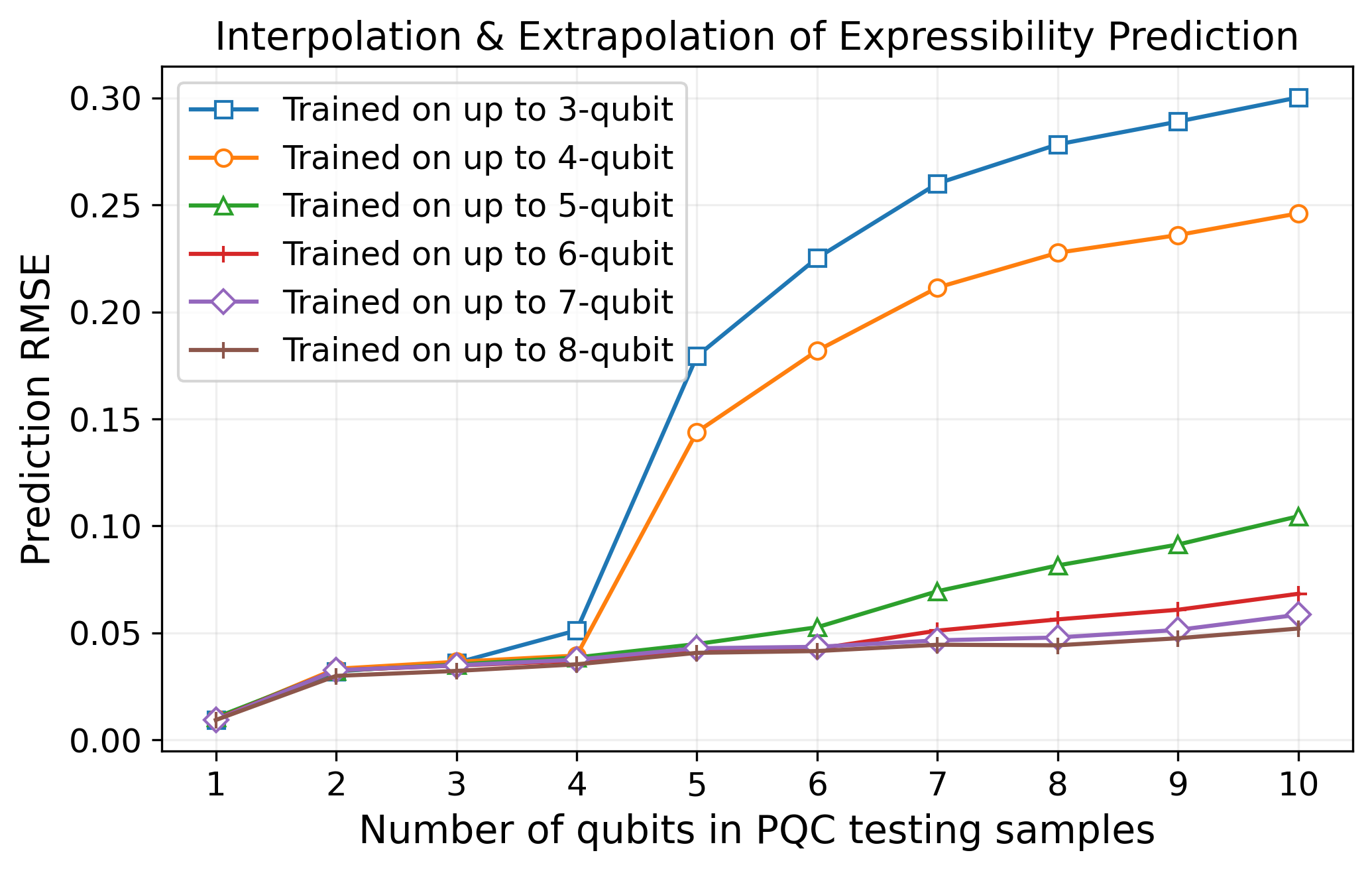}
    \caption{Interpolation \& Extrapolation prediction RMSE up to 10-qubit PQCs:
    Each line represents testing RMSE for a model trained on PQCs up to a fixed number of qubits (interpolation) and RMSE for that model on PQC sizes starting from the next value up to 10 qubits (extrapolation).
    Interpolation performs well for all models. For models trained only on PQCs on up to $4$ qubits, extrapolation has a very high RMSE. This decreases significantly with models trained on PQCs on up to $5$ or more qubits.}
    \label{fig:extrapolation-rmse}
\end{figure}

We also explore the model’s performance with higher qubit circuit sizes. In Figure~\ref{fig:extrapolation-rmse}, we present the interpolation and extrapolation of expressibility prediction results for qubits up to `10’. The plot illustrates the prediction RMSE for a model trained on data up to a specific qubit count. Each line includes results from `1’ to 10’ qubits, where the RMSE from `1’ to the current qubit count represents interpolation, and the RMSE from the next qubit to the 10th qubit represents extrapolation scenarios. 
For instance, the first experiment utilized training samples of PQCs on up to 3 qubits. It computed a testing RMSE of expressibility prediction for in-range (1--3~qubit(s)) and out-of-range (4--10~qubits) PQC sizes, respectively. 

Interpolation consistently exhibits good accuracy across the known qubit range, with consistently low RMSE. The average RMSE for interpolation across different qubit models is 0.05, indicating precise predictions within the trained domain. As we extend testing beyond the trained ranges, especially with lower qubit counts, extrapolation shows a notable increase in prediction RMSE. For instance, the RMSE for the `up to 3-qubit’ model spikes to 0.22 and 0.3 for extrapolating to 6 and 10-qubit circuit sizes, respectively, indicating significant deviation from expected values. Nevertheless, when the model is trained on up to higher qubit PQC sizes, such as the `up to 8-qubit’ model, it achieves an RMSE of 0.05 for predicting expressibility values for unseen `10’ qubit circuits. Notably, the plot indicates % that the model begins to exhibit 
good extrapolation accuracy already from the `up to 5-qubit’ model onwards.

\section{Related Works}
\label{sec:related}
\subsection{ML in Quantum Computing}
In recent years, the application of machine learning (ML) techniques in various aspects of quantum computing, including fidelity estimation, circuit optimization, error mitigation, and compilation, has gained considerable attention. 

\subsubsection{Fidelity Estimation}
Quantum fidelity is a similarity measure between an experimentally prepared mixed state and a target theoretical pure  state~\cite{jozsa1994fidelity}. Accurately estimating the fidelity of quantum states is essential for validating the reliability of a prepared quantum state. However, fidelity estimation is computationally expensive, presenting a challenge due to its inherent complexity. There has been a growing interest in employing machine learning-based and statistical methods to predict fidelity metric~\cite{liu2020reliability, yu2022statistical, zhang2021direct, Wang2022torchQuantum, saravanan2022data}. Yu~\etal~\cite{yu2022statistical} discusses statistical methods for fidelity estimation, and Zhang~\etal~\cite{zhang2021direct} proposed a machine learning-based direct fidelity estimation framework. Liu~\etal~\cite{liu2020reliability} propose a fidelity estimation model that employs polynomial fitting and shallow neural networks. Their methodology treats the quantum backend as a black box. Some recent works have focused on modeling quantum hardware noise using a deep learning-based data-driven approach. Wang~\etal~\cite{Wang2022torchQuantum} leveraged graph transformer networks, and Saravanan~\etal~\cite{saravanan2022data} utilized graph neural networks to predict the noise impact on quantum circuit fidelity by capturing complex circuit structure.

\subsubsection{Quantum Compilation}
Quantum compilation is the process of translating a high-level quantum algorithm or quantum circuit into a sequence of elementary quantum gates that can be executed on a specific quantum hardware platform. This translation involves several steps, including gate decomposition, gate optimization, and circuit mapping. ML approaches have been proposed in different quantum compilation phases, including gate decomposition~\cite{zhang2020topological}, circuit optimization
\cite{paler2023machine, fosel2021quantum, zlokapa2020deep}, and quantum circuit mapping~\cite{acampora2021deep,fan2022optimizing}. Moreover, Quetschlich~\etal employed reinforcement learning to develop optimized quantum circuit compilation flows~\cite{quetschlich2023compiler}. Recently, LeCompte~\etal effectively combined reinforcement learning with a Graph Neural Network (GNN)-based Q-network to allocate qubits for error reduction~\cite{lecompte2023machine}.

\subsubsection{Quantum Error Mitigation}
Quantum systems are inherently susceptible to various types of errors due to factors such as noise, decoherence, and imperfect operations. These errors can degrade the accuracy and reliability of quantum computations. Quantum error mitigation techniques are designed to reduce or compensate for errors that arise during quantum computations. Examples of such techniques include zero-noise extrapolation (ZNE), Probabilistic error cancellation~\cite{temme2017error}, and virtual distillation~\cite{huggins2021virtual}.
However, these techniques have considerable computational costs, including exponential sampling overheads. To resolve this, researchers attempted ML approaches for efficient quantum error mitigation~\cite{strikis2021learning,liao2023machine,kim2020quantum,sayar2022ssqem}. Strikis~\etal~\cite{strikis2021learning} used a learning-based protocol for improving probabilistic error cancellation. Moreover, Kim~\etal~\cite{kim2020quantum} used an artificial neural network with a shallow depth quantum circuit to estimate output probability adjustments. The semi-supervised error mitigation technique from~\cite{sayar2022ssqem} first learns the error types and then mitigates them using deep learning approaches. Recently, Liao~\etal~\cite{liao2023machine} employed various ML techniques to significantly reduce overheads while maintaining the accuracy of conventional error mitigation methods.

\subsection{Expressibility Prediction}
Expressibility as a metric has been defined in~Sim~\etal~\cite{sim2019expressibility} as KL divergence of the (estimated) fidelity distribution of a PQC from the uniform fidelity distribution of a Haar random unitary. The study employs a circuit set spanning various application domains and computes the expressibility of PQCs using the defined technique. To the best of our knowledge, no prior work has been done to predict the expressibility of PQCs using ML techniques. Therefore, in this work, we evaluate our expressibility prediction with computed expressibility values using the reference circuit set, and most commonly utilized hardware-efficient ansatz sets from IBM Qiskit.

\section{Conclusion}
\label{sec:conclusion}
We introduce a machine learning-based approach to estimate the expressibility of PQCs across diverse quantum machines, mitigating challenges associated with existing statistical estimation techniques. By conceptualizing PQCs as graphs and leveraging graph neural networks, our method efficiently evaluates PQCs by capturing intricate relationships between circuit architecture and expressibility. 
After training the GNN model with a diverse dataset of PQCs, our prediction model can significantly reduce the exponential computational cost associated with fidelity distribution estimation in expressibility computation.

Experimental evaluation on a comprehensive dataset from noiseless and noisy backends shows the predictive accuracy of the model, demonstrating a close alignment between predicted and actual expressibility values. 
Additionally, our model demonstrates strong extrapolation abilities by accurately predicting expressibility values for out-of-range qubit circuits. Furthermore, we observe the impact of noise on PQCs affecting the performance of variational quantum algorithms in different quantum backends. 

%In future work, we aim to expand the prediction model to accommodate PQCs with a more significant number of qubits and to generalize it for a broader range of quantum backends.

% Commented out for blind peer review
%\label{sec:conclusion}
\section*{Acknowledgements}
The research presented in this article was supported by the Laboratory Directed Research and Development program of Los Alamos National Laboratory under project number 20230049DR. 
Ms.\ Aktar is supported by a College of Engineering at New Mexico State University Graduate Fellowship. 
%Los Alamos National Laboratory is managed by Triad National Security, LLC, for the National Nuclear Security Administration of the U.S. DOE under contract 89233218CNA000001.
Los Alamos unlimited release report number: LA-UR-23-33850. 
%\vspace{-2mm}

%\bibliographystyle{IEEEtranS} %% for conference camera-ready version
\bibliographystyle{plainurl} %% for arXiv and for conference submissions
\bibliography{main-arxiv}

\end{document}